\documentclass[
 superscriptaddress,
 reprint,
 amsmath, amssymb,
 aps, preprintnumbers, nofootinbib, floatfix, longbibliography
]{revtex4-1}

\usepackage{graphicx}
\usepackage{dcolumn}
\usepackage{bm}
\usepackage{ulem}
\usepackage{color}
\usepackage{amssymb}
\usepackage{amsmath}
\usepackage{comment}
\usepackage{adjustbox}
\usepackage[colorlinks=true,allcolors=Blue,pdfusetitle]{hyperref}
\usepackage{rotating}
\usepackage{multirow}
\usepackage{graphicx}
\usepackage{lipsum}
\usepackage[dvipsnames]{xcolor}
\usepackage{svg}
\usepackage{soul}
\usepackage{esdiff}
\definecolor{Green}{RGB}{0, 128, 0}
\newcommand{\orcid}[1]{\href{https://orcid.org/#1}{\includegraphics[width=10pt]{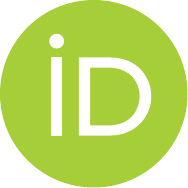}}}

\begin{document}
\preprint{N3AS-22-011}
\title{Exploiting stellar explosion induced by the QCD phase transition in large-scale neutrino detectors}

\author{Tetyana Pitik \orcid{0000-0002-9109-2451}}
\email{tetyana.pitik@nbi.ku.dk}
\affiliation{
Niels Bohr International Academy and DARK, Niels Bohr Institute, \\
University of Copenhagen, Blegdamsvej 17, 2100, Copenhagen, Denmark
}

\author{Daniel J. Heimsoth \orcid{0000-0002-5110-1704}}
\email{dheimsoth@wisc.edu}
\affiliation{
Department of Physics, University of Wisconsin--Madison,
Madison, Wisconsin 53706, USA}

\author{Anna M. Suliga \orcid{0000-0002-8354-012X}}
\email{asuliga@berkeley.edu}
\affiliation{
Department of Physics, University of California Berkeley, Berkeley, California 94720, USA}
\affiliation{
Department of Physics, University of Wisconsin--Madison,
Madison, Wisconsin 53706, USA}

\author{A. Baha Balantekin \orcid{0000-0002-2999-0111}}
\email{baha@physics.wisc.edu}
\affiliation{
Department of Physics, University of Wisconsin--Madison,
Madison, Wisconsin 53706, USA}

\date{November 4, 2022}

\begin{abstract}

The centers of the core-collapse supernovae are one of the densest environments in the Universe. Under such conditions, it is conceivable that a first-order phase transition from ordinary nuclear matter to the quark-gluon plasma occurs. This transition releases a large amount of latent heat that can drive a supernova explosion and may imprint a sharp signature in the neutrino signal. We show how this snap feature, if observed at large-scale neutrino detectors, can set competitive limits on the neutrino masses and assist the localization of the supernova via triangulation. The 95\%C.L. limit on the neutrino mass can reach  0.16~eV in Ice-Cube, 0.22~eV in Hyper-Kamiokande, and 0.58~eV in DUNE,  for a supernova at a distance of 10 kpc. For the same distance and in the most optimistic neutrino conversion case, the triangulation method can constrain the $1\sigma$ angular uncertainty of the supernova localization within $\sim 0.3^{\circ}-9.0^{\circ}$ in the considered pairs of the detectors, leading to an improvement up to an order of magnitude with respect to the often considered in the literature rise time of the neutronization burst.
\end{abstract}

\maketitle


\section{Introduction}
\label{sec:Introduction}

The detection of electron antineutrinos from SN~1987A~\cite{PhysRevLett.58.1490, Bionta:1987qt, ALEXEYEV1988209} was a significant milestone in multimessenger astronomy. It provided direct evidence that, in their latest stages of life, massive stars emit a considerable fraction of the gravitational binding energy in the form of neutrinos. 
Many aspects of core-collapse supernovae (CCSNe) are well investigated (for recent reviews see, e.g, Refs.~\cite{Janka:2017vlw, Burrows:2020qrp}).
During the infall phase each element of matter moves with constant entropy~\cite{Brown:1982cpw}. After reaching sufficiently high densities, electrons get captured on heavy nuclei and protons,  producing an initial sharp burst of electron neutrinos with typical luminosities of the order $\sim$ few $10^{53}$~erg~\cite{Bethe:1979zd, Fuller:1981mv}. This is followed by a smooth signal of neutrinos of all flavors created primarily by pair-production processes during the accretion and cooling of the newly formed protoneutron star~\cite{Mirizzi:2015eza, Horiuchi:2018ofe}.

There are several questions regarding CCSNe that remain unanswered. One unsettled crucial issue is the mechanism by which CCSNe explosions occur. The most widely explored hypothesis is the delayed neutrino heating mechanism~\cite{Colgate:1966ax, Bethe:1985sox}. 
This scenario, in a simplified fashion, can be explained in the following way: after the core bounce, the stalled shock wave is revived by an increase in thermal pressure due to reabsorption (neutrino heating) of a fraction of the neutrinos emitted (neutrino cooling) from the protoneutron star. In reality, the process of neutrino decoupling from the matter, which dictates the neutrino heating and cooling rates, is a complex phenomenon determined by the neutrino absorption, emission, and scattering reactions with the matter~\cite{Raffelt:2001kv, Fischer:2011cy}.
The neutrino heating aids in the development of hydrodynamical instabilities~\cite{Blondin:2002sm, Tamborra:2014aua, Couch:2014kza, Abdikamalov:2014oba} which in turn have been shown to produce explosions in many of the state-of-the-art three-dimensional supernova simulations. 
However, these models only explode for stars with masses below approximately $25~M_\odot$ and do not produce large explosion energies (hypernovae)~\cite{Galama:1998ea, Stanek:2003tw, Hjorth:2003jt, DellaValle:2006yp}.

Besides neutrino heating, other explosion mechanisms have also been proposed, e.g., magnetorotational supernova mechanism~\cite{LeBlanc:1970kg, Takiwaki:2007sf, Mosta:2015ucs, Obergaulinger:2020cqq, Matsumoto:2022hzg}, the acoustic mechanism~\cite{Burrows:2005dv, Gossan:2019uzp}, and sterile neutrino decays~\cite{Fuller:2008erj, Rembiasz:2018lok, Mastrototaro:2019vug} and conversions~\cite{Qian:1993dg, Sigl:1994hc, Hidaka:2006sg, Suliga:2020vpz}. One of the most notable of them is the quark-hadron phase transition (QHPT) mechanism~\cite{Sagert:2008ka, Fischer:2010wp, Fischer:2017lag, Zha:2021fbi, Fischer:2021tvv, Kuroda:2021eiv, Bauswein:2022vtq, Lin:2022lck, Jakobus:2022ucs} in which the stellar explosion is connected with a first-order phase transition from nuclear to quark matter.
The latent heat released during the transition can launch a second shock wave which can help drive the explosion. As this fast shock passes the neutrinosphere, a sharp burst containing mostly electron antineutrinos is released.
The detection of such a sharp feature in the neutrino signal would not only strongly support the occurrence of the phase transition in SN cores but also enable testing the physics affecting neutrino propagation.

In this work, we show that the detection of the QHPT peak in neutrino signal by the three existing and future large-scale detectors -- Ice-Cube~\cite{Halzen:2009sm}, Hyper-Kamiokande~\cite{Hyper-Kamiokande:2018ofw}, and DUNE~\cite{DUNE:2020ypp} -- can set competitive limits on neutrino masses and can greatly improve the precision of localization of the supernova~\cite{Dasgupta:2009yj}. Two effects make this feasible -- the sharpness and intensity of the QHPT peak. 

This paper is organized as follows; in Sec.~\ref{sec:QGPT-SN} we briefly describe the supernova explosion mechanism due to QHPT, show the nominal neutrino signal expected from the QHPT, and discuss the impact of neutrino conversions inside the star on the neutrino propagation through the supernova medium.
In Sec.~\ref{sec:Questions} we summarize the status of supernova triangulation and the limits on neutrino masses, and show how they can be improved by employing the rise time of the sharp signal feature.
Sections~\ref{sec:Detection} and~\ref{sec:Answers} describe the detectors employed in our work and the results obtained, respectively.
Finally, we summarize and discuss our results in Sec.~\ref{sec:Conclusions}.

\section{Neutrino signal from QCD Phase Transition Supernova}
\label{sec:QGPT-SN}

In this section, we review the supernova explosion mechanism due to a quark-hadron phase transition (Sec.~\ref{sec:QHPT-Supernova-Explosions-Mechanism}) and highlight how it compares to neutrino fluxes without such transition (Sec.~\ref{sec:Luminosity}). In short, the QHPT produces an additional narrow peak in the neutrino signal during the accretion phase, as we elaborate below.

\subsection{Quark-hadron phase transition as an engine for supernova explosion}
\label{sec:QHPT-Supernova-Explosions-Mechanism}

When a star with a mass $M \gtrsim 8$~M$_{\odot}$ has reached the end of its life and its core has attained sufficiently high densities, the photodisintegration of heavy nuclei as well as electron captures on nuclei and free protons reduce the pressure in the inner core and eventually lead to collapse on a dynamical timescale~\cite{Janka:2017vlw}.
Electron captures produce a huge amount of electron neutrinos that get trapped inside the core once the central density exceeds $\sim$ few $10^{11}$~g cm$^{-3}$~\cite{Bethe:1979zd}.

In the remaining collapse when nuclear saturation densities of about $2.3 \times 10^{14}~\mathrm{g}\;\mathrm{cm}^{-3}$ are reached, the collapse is halted by the repulsive short-range nuclear forces, leading to a stiffening of the equation of state (EoS) and a (first) core bounce.
A protoneutron star (PNS) is left at the center of the star and a shock wave is launched. As the shock crosses the neutrinosphere, the neutronization burst of electron neutrinos is released.
The neutrino losses and continuous energy loss due to the dissociation of in-falling of heavy nuclei from the outer core quickly cause the shock to stall~\cite{Bethe:1990mw,Woosley:2002zz}.

The following evolution of the EoS and shock wave determine whether the star explodes and what kind of a compact remnant is formed, a black hole (BH), a neutron star (NS), or another kind of compact star containing nonhadronic matter~\cite{Gerlach:1968zz, Alford:2013aca, Espino:2021adh}. Given that the state of matter at densities well exceeding the nuclear saturation density is highly uncertain, the question about the composition of the central object is not unsubstantiated.

The possibility of a phase transition between the hadronic and quark matter has been studied extensively in the context of CCSNe~\cite{Migdal:1979je, Takahara:1988yd, Gentile:1993ma, Pons:2001ar, Drago:2008tb, Nakazato:2008su, Sagert:2008ka, Fischer:2010wp, Fischer:2017lag, Zha:2021fbi, Fischer:2021tvv, Kuroda:2021eiv, Bauswein:2022vtq, Lin:2022lck, Jakobus:2022ucs}. In several cases, it has been found that close to saturation density, for high temperatures and low proton fractions, the phase transition can occur in the early postbounce phase of a CCSN. In such cases, the accelerated contraction of the PNS in the mixed quark-hadron phase may lead to development of a pure quark phase.
The associated stiffening of the EoS causes an abrupt halt of further contraction; a second strong shock wave forms, which can merge with the initial stalled bounce shock and trigger the SN explosion~\cite{Sagert:2008ka}.

The QHPT provides an alternative mechanism for a successful SN explosion. It has also been shown to create explosions for otherwise nonexploding SN models~\cite{Fischer:2017lag}.
Moreover, the propagation of the second shock across the neutrinospheres releases an additional $\sim $~ms burst of neutrinos. Such a burst is dominated by $\bar{\nu}_{e}$ because matter in PNS is neutron rich, and in the presence of a large abundance of positrons due to high temperatures achieved during the shock propagation positron capture dominates over electron capture, favoring production of $\bar{\nu}_{e}$ over $\nu_{e}$, $\nu_{x}$ and $\bar{\nu}_{x}$. We investigate the value of detecting this sharp feature in the SN neutrino signal in the remaining of our work.


\subsection{Numerical model}
\label{sec:Luminosity}

For our benchmark QHPT SN neutrino signal, we use the results from a one-dimensional general-relativistic radiation-hydrodynamical simulation with Boltzmann neutrino transport from Ref.~\cite{Fischer:2017lag} of a 50$M_\odot$ CCSN progenitor evolved using the AGILE-BOLTZTRAN tool~\cite{Liebendoerfer:2002xn, Mezzacappa:1993gn}; a hadron-quark hybrid EoS where deconfinement is taken into account via an effective string-flip potential was developed for the simulation~\cite{Fischer:2017lag} based on Refs.~\cite{Bastian:2020unt, Kaltenborn:2017hus}. The developed EOS is within approximately 95\%C.L.~region constraints from combining the gravitational wave observation of the binary neutron star merger and X-ray plus radio observations of the pulsars~\cite{Riley:2021pdl, Raaijmakers:2021uju}.
The used model results in a successful explosion with an energy approximately $2.3\times 10^{51}$~erg at the shock breakout.

\begin{figure*}[t]
\includegraphics[width=0.99\columnwidth]{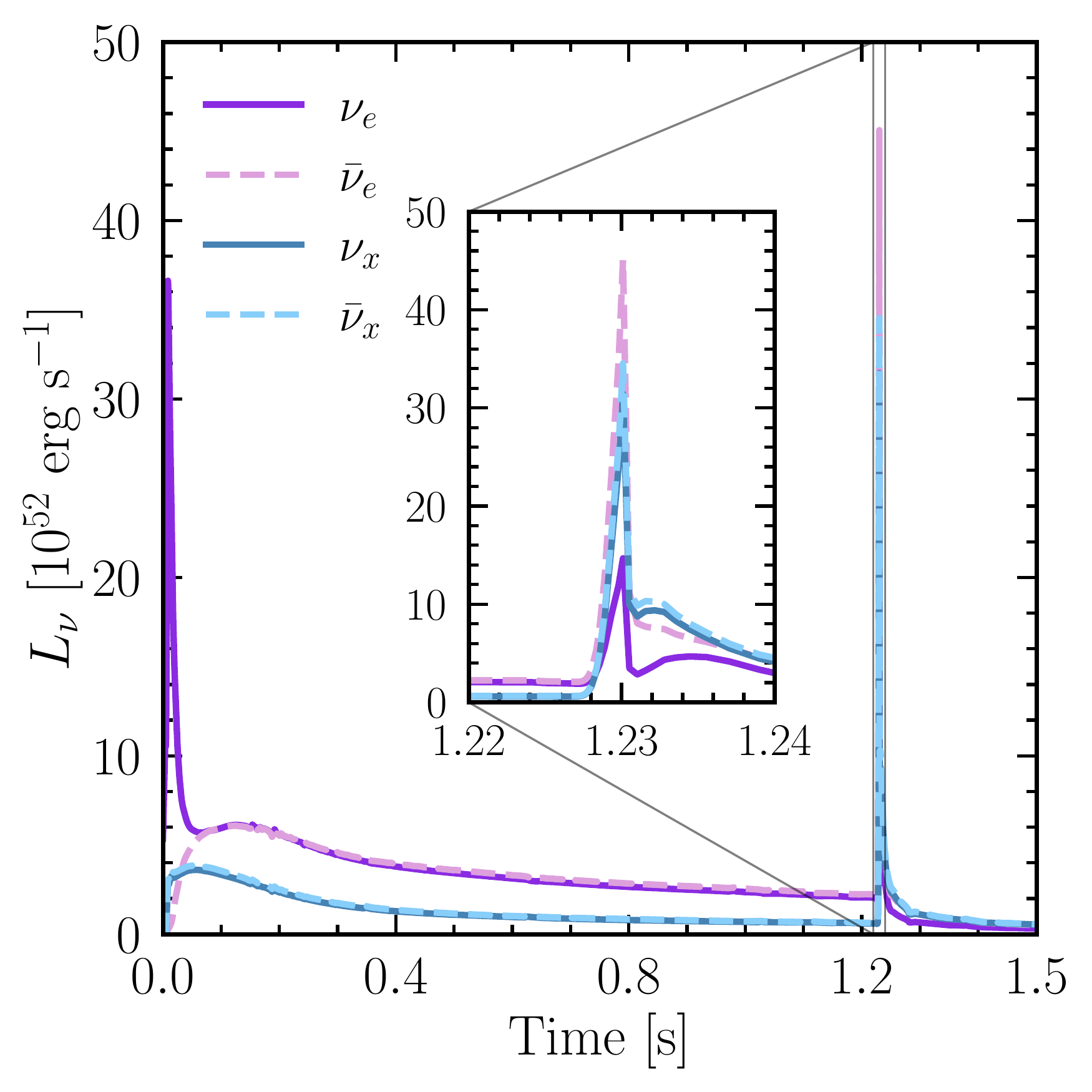} 
\includegraphics[width=0.99\columnwidth]{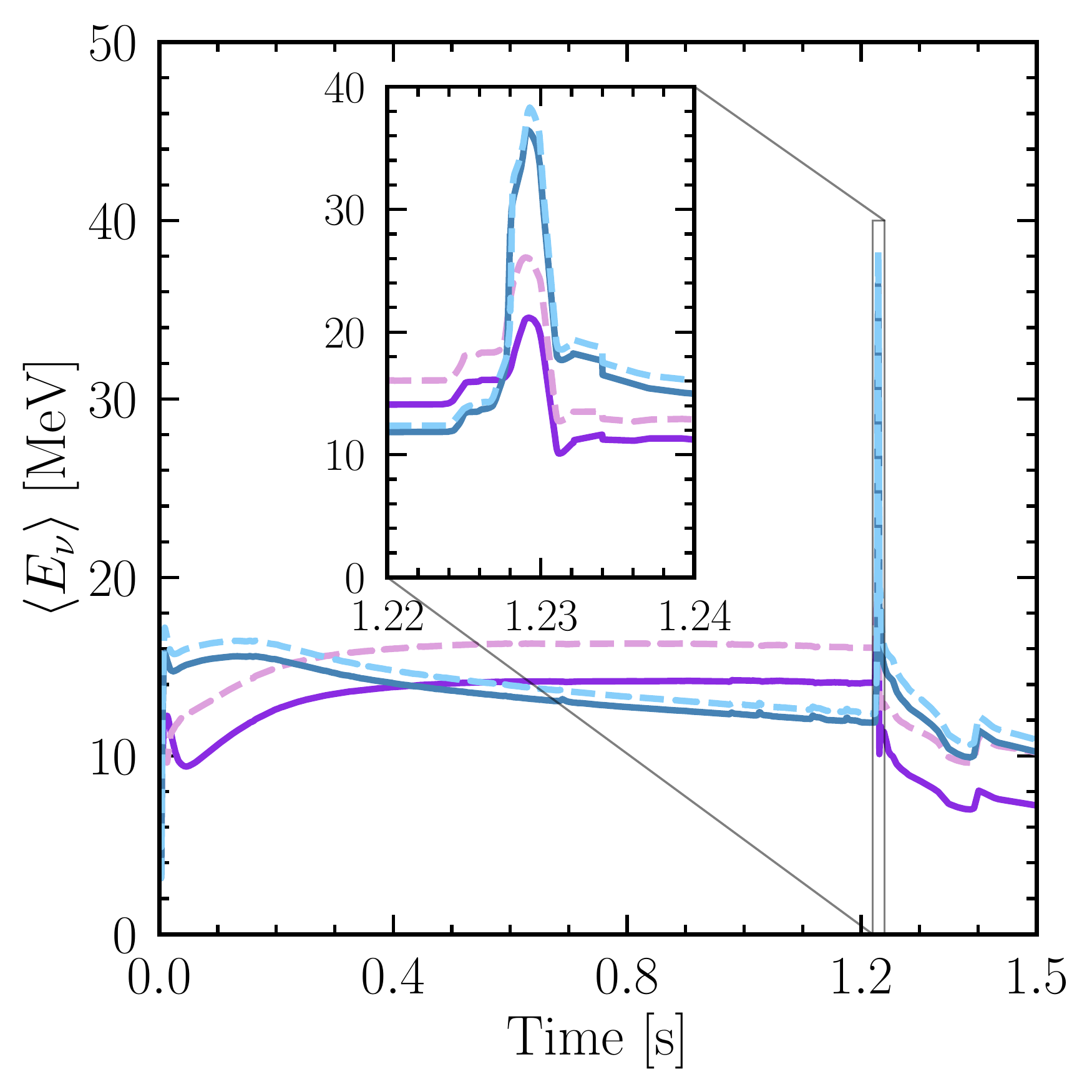}
\caption{Temporal evolution of the neutrino luminosities (left panel) and average neutrino energies (right panel) for $\nu_e, \bar\nu_e, \nu_x$, and $\bar\nu_x$ in the comoving frame of reference at 500 km, from a 50 $M_\odot$ CCSN progenitor with a quark-hadron phase transition~\cite{Fischer:2017lag}. As opposed to the neutronization burst at $\sim 8$~ms released after the initial core bounce, the second sharp peak at $\sim 1.23$~s is associated with the QCD phase transition.}
\label{Fig:luminosity_mean_energy}
\end{figure*}  

Figure~\ref{Fig:luminosity_mean_energy} shows the temporal evolution of neutrino luminosities (left panel) and average energies (right panel) from the employed QHPT SN model~\cite{Fischer:2017lag}. 
At early times one can see the neutronization burst of neutrinos released after the core bounce. At around $1.224$ s, after the creation of the pure quark matter phase, the second hydrodynamical shock takes over the standing bounce shock and triggers the onset of the SN explosion~\cite{Fischer:2017lag}. A few tens of milliseconds later (around $\sim 1.23$~s) the shock reaches the neutrinosphere and a second millisecond burst of $\bar{\nu}_{e}$ is released. Comparing its duration to the neutronization burst, the QHPT burst is much sharper~\cite{Sagert:2008ka, Fischer:2010wp, Fischer:2017lag}, with the width related to the energetics of the second shock.
In addition, as the QHPT shock travels through the neutrinospheres and heats up the matter, the mean neutrino energies increase sharply analogously to the luminosities.

\subsection{Neutrino spectra}
\label{sec:Neutrino-spectra}

To construct the neutrino fluxes from the QHPT SN, we assume that the outgoing neutrino spectra obey Fermi-Dirac distributions with negligible chemical potential, since the amount of spectral pinching in the signal after the neutronization phase is expected to be small as compared to the neutronization burst phase~\cite{Janka:2017vlw}; this is also true for the QHPT CCSN model we use, where the pinching parameter ($\alpha$) during the $\bar\nu_e$ burst varies between 2 and 4. 
These distributions can be easily recovered using the parametrization form~\cite{Keil:2002in, Keil:2003sw, Tamborra:2012ac},
\begin{equation}
\label{eq:energy-distribiution}
\begin{split}
	\varphi_{\nu_\beta}(E_{\nu}, t) & = \xi_{\nu_\beta}(t) \left( \frac{E_{\nu}}{\langle E_{\nu_\beta}(t)\rangle}\right)^{\alpha_{\nu_\beta}(t)} \\ & \times \exp \left(-\frac{(\alpha_{\nu_\beta}(t)+1)E_{\nu}}{\langle E_{\nu_\beta}(t)\rangle}\right) \ ,
\end{split}
\end{equation}
where $E_{\nu}$ is the neutrino energy, $\langle E_{\nu_\beta}(t)\rangle$ is the average neutrino energy, and the normalization factor $\xi_{\nu_\beta}(t) = (\int dE_{\nu} \varphi_{\nu_\beta}(E_{\nu}, t))^{-1}$. The pinching parameter $\alpha_{i}(t)$ for the Fermi-Dirac distribution is $\alpha_{i}(t) \equiv \alpha = 2.3$.
Finally, the differential flux of neutrinos ($\nu_\beta = \{\nu_{e},\bar{\nu}_{e},\nu_{x},\bar{\nu}_{x} \}$, where $\nu_x$ indicates $\nu_\mu$ or $\nu_\tau$, as to first order they are indistinguishable for the purposes of supernova modeling) can be obtained as follows:
\begin{equation}
\label{eq:flux}
	F_{\nu_\beta} (E_{\nu}, t) = \frac{L_{\nu_\beta}(t)}{\langle E_{\nu_\beta}(t)\rangle}\frac{\varphi_{\nu_\beta}(E_{\nu}, t)}{4\pi D^2} \ ,
\end{equation}
where $L_{\nu_\beta} (t)$ is the luminosity of neutrino flavor $\nu_{\beta}$, and $D$ is the distance to the SN.

\subsection{Neutrino flavor conversion in dense medium}
\label{sec:conversions-MSW}

While propagating through the SN medium, neutrinos undergo flavor conversions due to their self-interactions and coherent forward scattering of the medium particles. The former is a highly nonlinear phenomenon subject to numerous ongoing studies aiming to understand its impact on the neutrino flavor evolution~(see, e.g., Refs.~\cite{Duan:2010bg, Chakraborty:2016yeg, Tamborra:2020cul, Patwardhan:2021rej} and references therein). 
The latter effect leads to the well-known Mikheyev-Smirnov-Wolfenstein (MSW) resonance effects, experimentally observed for solar neutrinos~\cite{Wolfenstein:1977ue, Mikheev:1986if}. As a result of the MSW conversions, neutrino eigenstates change their flavor reflecting the changes in the density of electrons they traveled through. 

In order to take into account all the uncertainties related to the flavor transformation in the dense stellar environment, in this work we consider two extreme conversion scenarios, i.e., the no conversion case and full conversion between $\nu_e$ and $\nu_x$ ($\bar\nu_e$ and $\bar\nu_x$). In the case of electron neutrino detectors, this should coincide with the maximal variation in the uncertainty on the parameters of interest. For electron antineutrino detectors, however, that might not be the case, as although the luminosity of $\bar{\nu}_{e}$ is larger than the $\bar{\nu}_x$ one, the mean energies display the opposite behavior (see Fig.~\ref{Fig:luminosity_mean_energy}). In Sec.~\ref{sec:Answers} we show that for electron antineutrino detectors, the no conversion scenario yields the best results.

\section{EXPLOITING THE SHARP FEATURE OF THE NEUTRINO SIGNAL}
\label{sec:Questions}

In this section, we outline how the observation of the QHPT peak in the neutrino signal can be used to improve the localization of the supernova in the sky (Sec.~\ref{sec:triangulation}) and constrain the absolute neutrino mass (Sec.~\ref{sec:neutrino-mass}).

\subsection{Triangulation}
\label{sec:triangulation}

Since neutrinos interact weakly with matter, they escape almost unimpeded from the extremely dense regions of an exploding SN, making them the first messengers to inform us of the occurrence of the gravitational collapse. Due to their frequent interactions with the star's envelope, photons emerge more slowly with an  hours to days delay compared to neutrinos~\cite{Nakamura:2016kkl}.
In addition, the electromagnetic radiation might be obscured by dust surrounding the star as well as the interstellar medium or be absent for stars directly forming BHs. In such cases observing neutrinos may be the only way to obtain the information about the SN, apart from the possible simultaneous detection of gravitational waves~\cite{Antonioli:2004zb, Pagliaroli:2009qy, Warren:2019lgb, Pajkos:2020nti, Halim:2021yqa}.  Any prompt information about the SN direction represents a valuable tool to alert astronomers and allow for a rapid multimessenger follow-up.

Two types of localization techniques have been considered for SN neutrinos. The first one uses an individual detector for pointing and is based on angular distributions of neutrino-matter interaction products, which can be correlated with the neutrino direction. In this case, a single experiment can independently determine the SN direction and its uncertainty. One of the most considered channels is neutrino elastic scattering on electrons, which is a highly anisotropic interaction~\cite{Beacom:1998fj, Tomas:2003xn}.
The outgoing electron is scattered in the direction of the incoming neutrino momentum, allowing detectors with sufficient spatial resolution to reconstruct the direction of the neutrino source. As a reference, pointing with elastic scattering in Hyper-Kamiokande may reach precision of 1$^{\circ}$-1.3$^{\circ}$~\cite{Hyper-Kamiokande:2018ofw}.

In the context of multimessenger astronomy, it is always a good practice to use as much information as possible.
In this regard, the complementary method that has been considered is triangulation~\cite{Beacom:1998fj, Muhlbeier:2013gwa, Brdar:2018zds, Hansen:2019giq, Linzer:2019, Coleiro:2020vyj, Sarfati:2022}, which is based on measuring the arrival time delay between pairs of widely separated detectors around the globe. Such technique requires an immediate sharing of data among the different experiments.

Past works on this subject have focused on two features of the neutrino signal, the neutronization burst~\cite{Beacom:1998fj, Scholberg:2009jr, Muhlbeier:2013gwa, Brdar:2018zds, Hansen:2019giq, Linzer:2019, Coleiro:2020vyj} and the signal demise at BH formation~\cite{Muhlbeier:2013gwa, Brdar:2018zds, Hansen:2019giq, Sarfati:2022}. The key property shared by these features is their swift change in the number of neutrinos. During the neutronization burst, the electron neutrino signal sharply increases over a tens of ms timescale, whereas a rapid cutoff of neutrinos in the span of few ms follows the BH formation.

Our work seeks to test the feasibility of using the neutrino flux peak caused by the QHPT to time the neutrino signal in an effort to triangulate a SN. The advantage of this approach over using the neutronization burst is that the QHPT peak has a much shorter rise time, on the order of ms. It also does not rely on the SN creating a BH to produce a neutrino cutoff, which occurs an estimated 10\%-40\% of the time based on theory~\cite{Ertl:2016, Couch:2020, Sukhbold:2016} and observations~\cite{Adams:2017a, Adams:2017b, Neustadt:2021} and may suffer from gravitational redshift softening the cutoff for black hole formation~\cite{Wang:2021elf}.

\subsection{Absolute neutrino mass}
\label{sec:neutrino-mass}

While the observation of neutrino oscillations confirmed the existence of nonzero neutrino masses~\cite{Super-Kamiokande:1998kpq, Hirata:1992ku, Bellerive:2016byv}, from the oscillation experiments alone we can only learn about the squared mass differences. The measured squared mass splittings, however, do set the lower limit on neutrino masses assuming that the lightest neutrino mass state is massless. In the case of normal ordering (NO) it is $\sum_i m_{\nu,i} > 0.06~\mathrm{eV}$, whereas for the inverted ordering (IO) $\sum_i m_{\nu,i} > 0.1~\mathrm{eV}$, where $i=1,2,3$ denotes the active neutrino mass state index.

The strongest upper limit on the sum of the neutrino masses comes from cosmological observations and is approximately $\sum_i m_{\nu,i} \lesssim 0.1~\mathrm{eV}$~\cite{Zyla:2020zbs}. While it may be tempting to consider this value as a strong hint for NO, these limits vary with various assumptions about the employed statistical procedures (see, e.g., Refs.~\cite{Gariazzo:2018pei, Jimenez:2022dkn, Chudaykin:2022rnl, Gariazzo:2022ahe}). 
Terrestrial experiments can also set upper limits on the neutrino mass based on energy and momentum conservation. The most stringent limit comes from the Karlsruhe Tritium Neutrino experiment (KATRIN)~\cite{KATRIN:2005fny}, which looks at the end point of the electron spectrum from beta decay of tritium, and is $m_{\nu_e} = \sqrt{\sum_i |U_{ei}|^2 m_{\nu,i}^2} < 0.8$~eV~\cite{KATRIN:2021uub}.

The existence of nonzero neutrino masses is not without consequences for neutrinos produced in CCSNe. It not only facilitates neutrino conversions but also introduces energy-dependent time delays in the neutrino flux arriving at the Earth~\cite{Zatsepin:1968kt}. The delay of a neutrino mass state ($m_{\nu}$) with energy ($E_{\nu}$) can be calculated as
\begin{equation}
\label{eq:time-delay}
\Delta t \approx 5.15 \left(\frac{D}{10~\mathrm{kpc}}\right) \left(\frac{m_\nu}{1~\mathrm{eV}}\right)^2 \left(\frac{10~\mathrm{MeV}}{E_\nu}\right)^2 \:\mathrm{ms} \ ,
\end{equation}
where $D$ is the distance to the SN.
The detection of a neutrino signal from a nearby SN can then be used to set limits on the neutrino masses~\cite{Zatsepin:1968kt}. The observation of neutrinos from SN~1987A set the mass limit of $m_{\nu} \lesssim 5.8~\mathrm{eV}$ at 95\%~C.L~\cite{Loredo:2001rx, Pagliaroli:2010ik}, where all three active neutrino mass states are assumed to have the same mass $m_\nu$ because the mass squared differences are much smaller than this limit.

The limits coming from a detection of neutrinos from a future CCSN calculated with different techniques (e.g., functional fit to the signal or timing first neutrino event with the gravitational wave signal) and detectors vary between approximately $m_{\nu} \lesssim 0.5-1.5~\mathrm{eV}$~\cite{Nardi:2003pr, Nardi:2004zg, Pagliaroli:2010ik, Lu:2014zma, Rossi-Torres:2015rla, Hansen:2019giq, Pompa:2022cxc} for the benchmark distance to the SN of 10~kpc. The limit is expected to be significantly more stringent than the one from SN~1987A neutrinos, mostly thanks to improvements in the detection of MeV-energy neutrinos from SN~\cite{Scholberg:2012id, SNEWS:2020tbu}.

In addition, using the sharp end of the neutrino signal in case of BH formation sets a limit on the neutrino mass of $m_{\nu} \lesssim 0.3-1.8~\mathrm{eV}$, again depending not only on the type and volume of the considered detectors~\cite{Beacom:2000ng, Hansen:2019giq} but also on the decay length of the signal. The latter is determined by the environment where the BH formation occurs. In cases where the accretion flow cannot be neglected~\cite{Gullin:2021hfv} it prolongs the neutrino signal causing the obtained mass limit to weaken.

In this work, we investigate the feasibility of employing the neutrino burst from QHPT SN in setting the upper limits on the absolute neutrino mass.
Given the sharpness of the QHPT peak, we expect that the observation of such signal in large-scale neutrino detectors will result in a more stringent limit on the neutrino mass than the one obtained from the detection of the neutronization burst.

We note that there also have been works investigating the possibility of determining the neutrino mass ordering with the neutronization rise time~\cite{Serpico:2011ir, Brdar:2022vfr} assuming only the impact of neutrino interactions with matter on the conversion probabilities. Since, however, we take a conservative approach with respect to the impact of neutrino-neutrino interactions on the neutrino flavor evolution inside the supernova core (see Sec.~\ref{sec:conversions-MSW}), we are not exploring this scenario.

\section{Detection of the Neutrino Burst}
\label{sec:Detection}

In order to resolve the sharp peak due to QHPT, we employ three of the largest existing and upcoming neutrino experiments: Hyper-Kamiokande (HK), Ice-Cube (IC), and the Deep Underground Neutrino Experiment (DUNE), which are described shortly in Secs.~\ref{sec:HK} - \ref{sec:DUNE}; the characteristics and angular coordinates for the three considered detectors are summarized in Table~\ref{table:detectors}.

\begin{figure}[t]
\includegraphics[width=1\columnwidth]{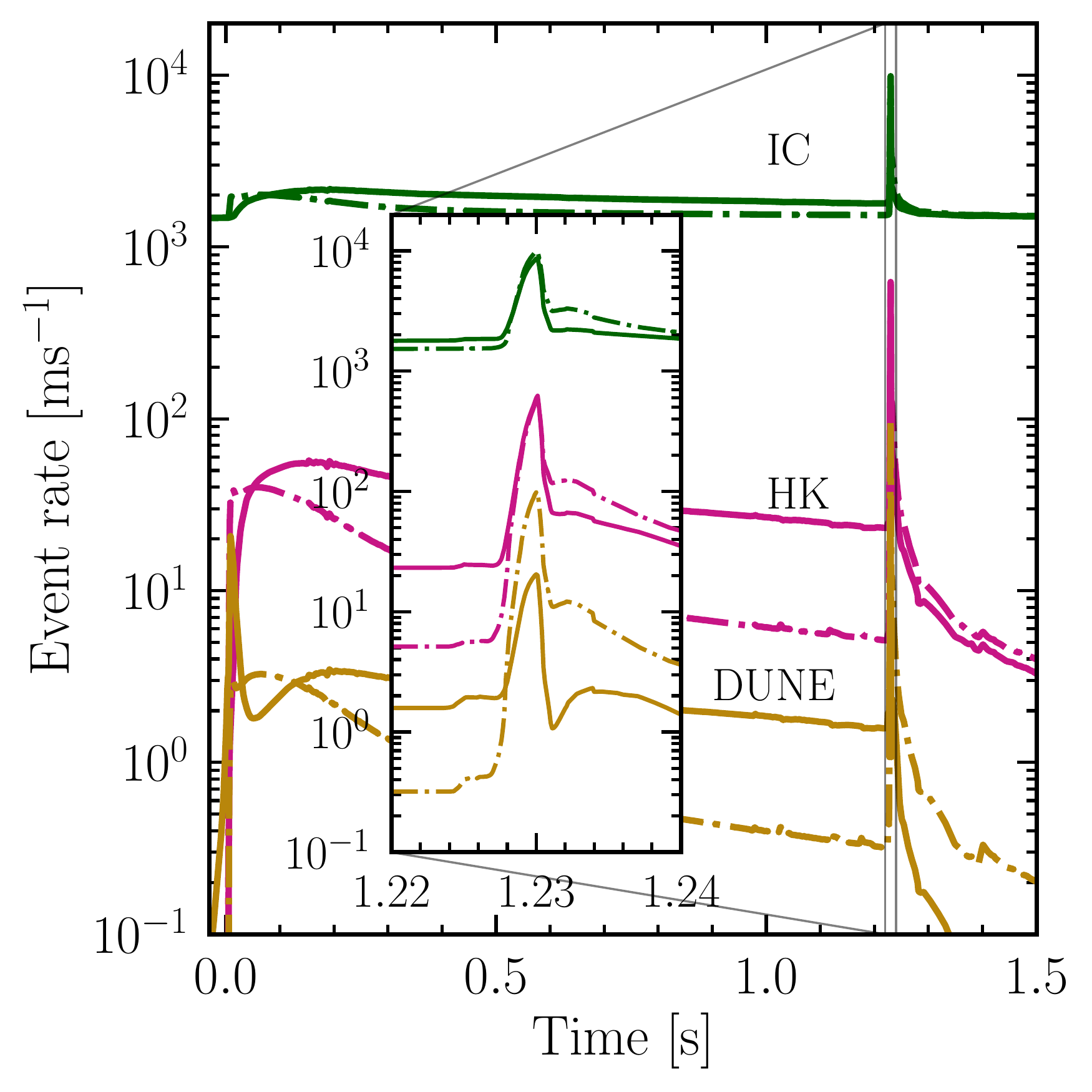} 
\caption{Event rates at the three detectors employed in this work HK, IC, and DUNE for the two considered conversion scenarios; no conversions (solid lines), full conversion (dash-dotted line). The event rate in the IC detector is the highest among the three detectors due to its largest detector volume and the smallest energy threshold.}
\label{fig:event-rate}
\end{figure} 

\begin{table*}[t]
	\caption{\label{table:detectors} Main characteristics of the detectors employed in our work: HK~\cite{Hyper-Kamiokande:2018ofw}, IC~\cite{Halzen:2009sm, IceCube:2011cwc}, and DUNE~\cite{DUNE:2020ypp, DUNE:2020zfm}.}
	\begin{center}
	\begin{adjustbox}{width=1\textwidth}
	\begin{tabular}{c|c|c|c|c|c|c|c}
		\toprule
		{Detector} & {Interaction} &{ Fiducial mass} &{Number of} & { Energy threshold}& {Latitude} & {Longitude} & {Distance between detectors}\\
		{} & {channel} &{[kt]} &{targets} & {[MeV]}& {[deg]} & {[deg]} & {[km]}\\		
		\hline
		IC &$\bar{\nu}_{e}+p\rightarrow e^{+}+n$ & $3500$&$6.18\times 10^{37}$ &$2.08$&-90&0 & D$_{\rm{IC-HK}}=11374$ \\
		HK & $\bar{\nu}_{e}+p\rightarrow e^{+}+n$ & $217$ & $1.45\times 10^{34}$ & $5$ &$36.4$ & $137.3$ & D$_{\rm{HK-DUNE}}=8369$ \\
		DUNE & $\nu_{e}+\,^{40}$Ar $\rightarrow e^- + ^{40}\mathrm{K}^*$ & $40$ & $6.02\times 10^{32}$  & $5$ &$44.4$ & $-103.8$& D$_{\rm{DUNE-IC}}=11746$ \\
		\toprule
	\end{tabular}
	\end{adjustbox}
	\end{center}
\end{table*}

In each of the three considered detectors, the energy-integrated event rate from the interaction channel $i$ can be calculated with the following formula:
\begin{equation}
\label{eq:event-rate}
R(t) =  N_{t} \int_{E_{\nu}^\mathrm{min}}^{\infty} dE_{\nu} \int_{E_\mathrm{th}}^{E_\mathrm{max}} dE \; \varepsilon \sigma_{i}(E,E_{\nu}) \; F_{\nu_\beta}(E_{\nu}, t) \ ,
\end{equation}
where $\varepsilon$ is the efficiency of the detector, which reflects what percentage of the detected signal can be directly translated into the event rate, $N_t$ is the number of targets in the detector volume, $\sigma_{i}(E,E_{\nu}) $ is the differential cross section for the interaction $i$, $E_\mathrm{th}$ is the threshold energy for the detector with $E_{\nu}^\mathrm{min}$ the corresponding neutrino energy, and $E_\mathrm{max}$ is the maximum energy that can be transferred to the detectable particle.

\subsection{Hyper-Kamiokande}
\label{sec:HK}

\begin{figure}[h!]
	\includegraphics[width=0.94\columnwidth]{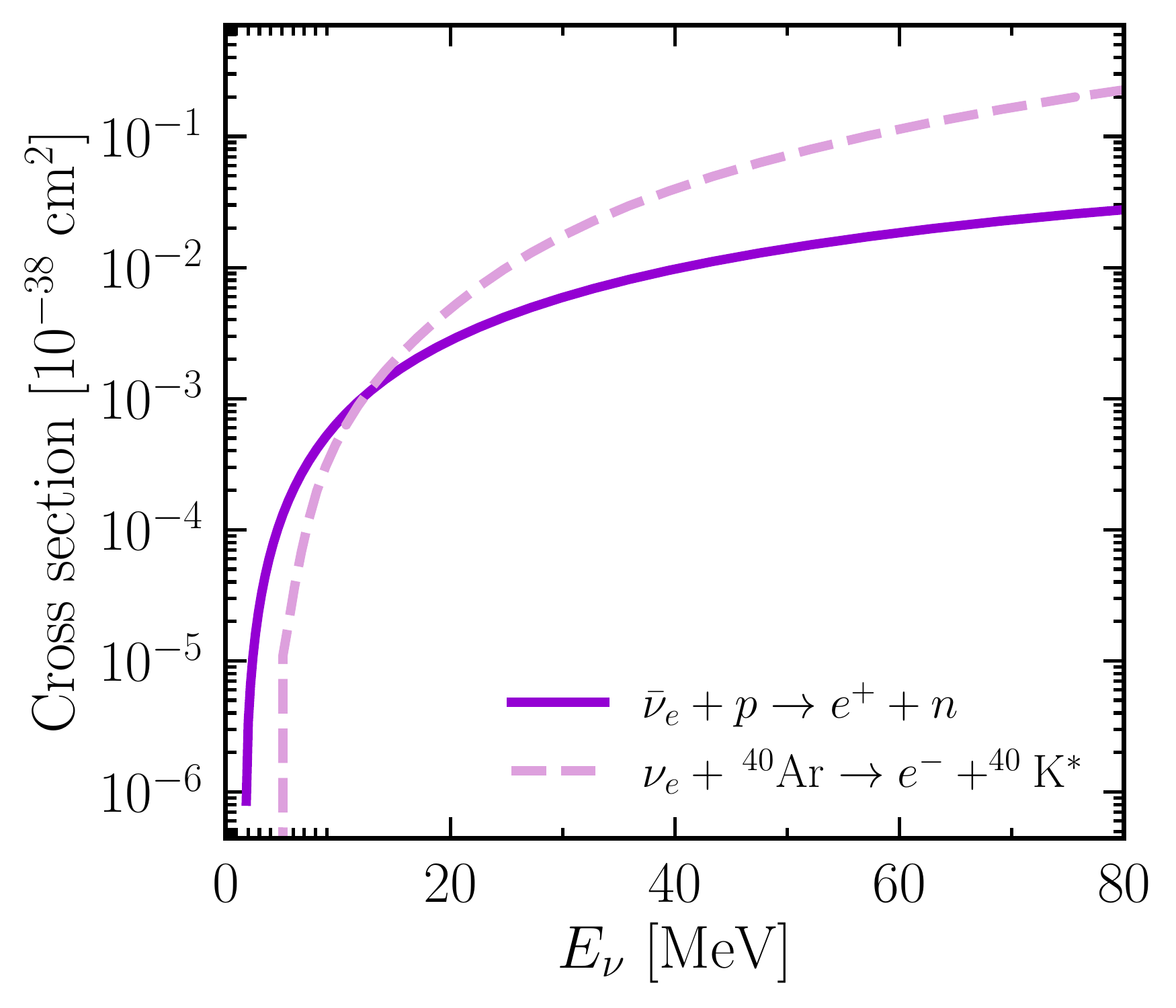}
	\caption{The cross sections as a function of the neutrino energy for the IBD scattering of $\bar{\nu}_{e}$ (purple solid line), and for the CC $\nu_{e}$ scattering on Argon (light violet dashed line). }
	\label{Fig:Cross section}
\end{figure}

HK~\cite{Hyper-Kamiokande:2018ofw}, which is currently under construction in Japan, is a 217~kt water Cherenkov detector. The main detection channel for MeV-energy-range SN neutrinos is inverse beta decay (IBD): $\bar{\nu}_{e}+p\rightarrow e^{+}+n$. This detection channel relies on observing the Cherenkov radiation emitted by the positron to identify the individual neutrino IBD reaction. 
In our work we neglect the small contribution coming from other reactions such as elastic scattering on electrons or neutral current neutrino-oxygen scattering, and focus only on the IBD channel. While it makes our analysis more conservative, it does not significantly affect the results.
We compute the HK event rate using Eq.~\ref{eq:event-rate} with the cross section for IBD~\cite{Strumia:2003zx, Ricciardi:2022pru} shown in Fig.~\ref{Fig:Cross section} as a function of the neutrino energy, positron energy threshold 5~MeV, and $\varepsilon = \,$100\% efficiency~\cite{Hyper-Kamiokande:2018ofw}. The temporal resolution of HK is expected to be of the order of ns~\cite{Hyper-Kamiokande:2018ofw}. According to the Ref.~\cite{Linzer:2019} the background rate in 217~kton HK can be as large as $10^{-4}~\mathrm{ms}^{-1}$; however, since we are focusing on a short time window for the QHPT peak (few ms) and do not look for time differences between single events, we neglect any contribution from background events in our analysis.

\subsection{Ice-Cube}
\label{sec:IC}

The IC neutrino observatory is an ice Cherenkov neutrino detector located at the South Pole~\cite{Halzen:2010yj}. It consists of 5160 digital optical modules (DOMs), each of a volume  capable of observing MeV-energy-range neutrinos from SN through IBD~\cite{IceCube:2011cwc}. While in this energy range IC cannot resolve the energy of individual neutrinos, the overall background rate will be enhanced signifying the SN signal. The temporal resolution for IC can reach down to few ns~\cite{IC-thesis:2015}. 

To obtain the photoelectron counts in the IC detector we follow the procedure described in Ref.~\cite{IceCube:2011cwc}. The total IBD event rate in the volume of the detector can be estimated with
\begin{equation}
\label{eq:event-rate-IC}
    R_{\rm{IC}}(t)  = \varepsilon \tilde{R}_{\rm{IC}}(t) + \mathrm{B}_{\mathrm{IC}} = \frac{0.87}{1 + \tilde{R}_{\rm{IC}}(t) \cdot \tau} \tilde{R}_{\rm{IC}}(t) + \mathrm{B}_{\mathrm{IC}} \, 
\end{equation}
where $\epsilon$ is the dead-time efficiency. With the dead time $\tau = 250~\mu\mathrm{s}$, the background noise can be reduced to the level of $\mathrm{B}_{\mathrm{IC}} = 286~\mathrm{Hz}$~\cite{IceCube:2011cwc}.
The IBD rate without accounting for background noise and the dead-time correction [which serves as the energy dependent efficiency ($\varepsilon$) of the detector], is given by~\cite{IceCube:2011cwc}
\begin{equation}
\label{eq:event-rate-IC-2}
\begin{split}
\tilde{R}_{\rm{IC}}(t) & = N_\mathrm{DOM} n_{p} \int_{E_{\nu}^\mathrm{min}}^{\infty} dE_{\nu} \times \\ 
&\int_{E_{e}^{\mathrm{min}}}^{E_{e}^{\mathrm{max}}} dE_{e} \; V_\mathrm{eff} N_\gamma(E_{e}) \; \sigma_{\rm{IBD}}(E_{e},E_{\nu}) \; F_{\bar{\nu}_{e}}(E_{\nu}, t) \ ,
\end{split}
\end{equation}
where $N_\mathrm{DOM}$ is the number of DOMs, $n_p \approx 6 \times 10^{22}~\mathrm{cm}^{-3}$ is the density of targets in the ice, $V_\mathrm{eff} \approx 1.63 \times 10^5~\mathrm{cm}^3$ is the effective volume for a single photon, $N_\gamma (E_{e}) \approx 178 (E_{e}/\rm{MeV)}$ is the number of Cherenkov radiated photons by an electron with energy $E_{e}$, and $\sigma_{\rm{IBD}}(E_{e},E_{\nu})$ is the differential IBD cross section.

\subsection{DUNE}
\label{sec:DUNE}

DUNE near and far detectors will be constructed in Batavia and South Dakota, United States~\cite{DUNE:2020ypp}, respectively. For the purpose of SN neutrino detection, we consider the far detector which is planned to be a 40~kton liquid argon time projection chamber. 
Liquid argon has a particular sensitivity to the $\nu_{e}$ component of a SN neutrino burst, since the main detection channel for MeV-energy neutrinos is $\nu_{e}+^{40}\,\!\rm{Ar}\rightarrow e^{-}+^{40}\,\!\rm{K}^{\ast}$. The observable for this channel is the $e^{-}$ plus deexcitation products from the excited $^{40}\,\!\rm{K}^{\ast}$ final state. We use the total cross section for this charged-current reaction as reported in Refs.~\cite{DUNE:2020zfm,DUNE:2020ypp}, obtained by simulating the interaction of MeV neutrinos with nuclei in liquid argon through the event generator MARLEY, see Fig.~\ref{Fig:Cross section}. Comparable cross section has been calculated in Ref.~\cite{Suzuki:2012ds, ToshioSuzukiprivate}. The energy-dependent detector efficiency $\varepsilon$ together with the energy threshold are taken from Ref.~\cite{DUNE:2020zfm}.
DUNE can also detect $\bar\nu_e$ and $\nu_x$ by the charged-current and neutral current interactions on argon. However, we neglect these contributions as they are suppressed by at least an order of magnitude compared to $\nu_e$ induced event rate~\cite{DUNE:2020zfm}.
The background rate in DUNE originating from solar neutrinos~\cite{Capozzi:2018dat}, fast neutrons~\cite{Li:2020ujl} and spallation~\cite{Zhu:2018rwc} is expected to be approximately $4\times 10^{-3}~\mathrm{ms}^{-1}$; therefore, similarly as for HK, we neglect the contribution from the background in our analysis.

\subsection{Event rates}
\label{sec:event-rates}

Figure~\ref{fig:event-rate} shows the total neutrino event rate in the three considered detectors IC (green lines), HK (pink lines), and DUNE (yellow lines) from the QHPT SN occurring at a distance of 10~kpc from the Earth for the no conversion (solid lines) and full conversion (dash-dotted lines) cases.  

The highest event rate for DUNE is for the case of full conversion due to a higher luminosity and average neutrino energy for $\nu_{x}$ than $\nu_{e}$ (see Fig.~\ref{Fig:luminosity_mean_energy}).
For water Cherenkov detectors, the event rates during the QHPT peak in both conversion scenarios are comparable; the larger $\bar{\nu}_{e}$ luminosity compared to $\bar{\nu}_{x}$ is balanced by smaller average energy of $\bar\nu_e$ versus $\bar{\nu}_{x}$ at the peak (see Fig.~\ref{Fig:luminosity_mean_energy}). During the pre-QHPT-burst phase the difference in the event rates for the two conversion scenarios is more prominent.

The event rate in DUNE is low compared to HK and IC due to the smaller detector volume and lower efficiency, especially for neutrino energies between 5-12~MeV. It is also worth noting that the event rate in the IC detector during the QHPT peak dominates the background noise, which is not the case for SN models without QHPT~\cite{IceCube:2011cwc}.

In all three detectors the QHPT neutrino burst is distinguishable from a neutrino signal without such feature at approximately more than $3\sigma$ level for both conversion scenarios until a distance of $\sim 20$~kpc. Therefore, in Sec.~\ref{sec:Answers} we limit our analyses to that distance.

\section{METHODS AND RESULTS}
\label{sec:Answers}

In this section we present our results on SN triangulation and limits on absolute neutrino mass from the detection of the sharp QHPT peak in the SN neutrino signal. In Sec.~\ref{sec:timing} we summarize the method to determine the timing of the neutrino signal. Later in Sec.~\ref{sec:results-triangulation}--\ref{sec:combined-localization} we show our findings for SN triangulation, while in Sec.~\ref{sec:results-neutrino-mass} we calculate the sensitivity to the absolute neutrino mass.

\subsection{Timing the neutrino signal}
\label{sec:timing}

\begin{figure}[h!]
	\includegraphics[width=0.94\columnwidth]{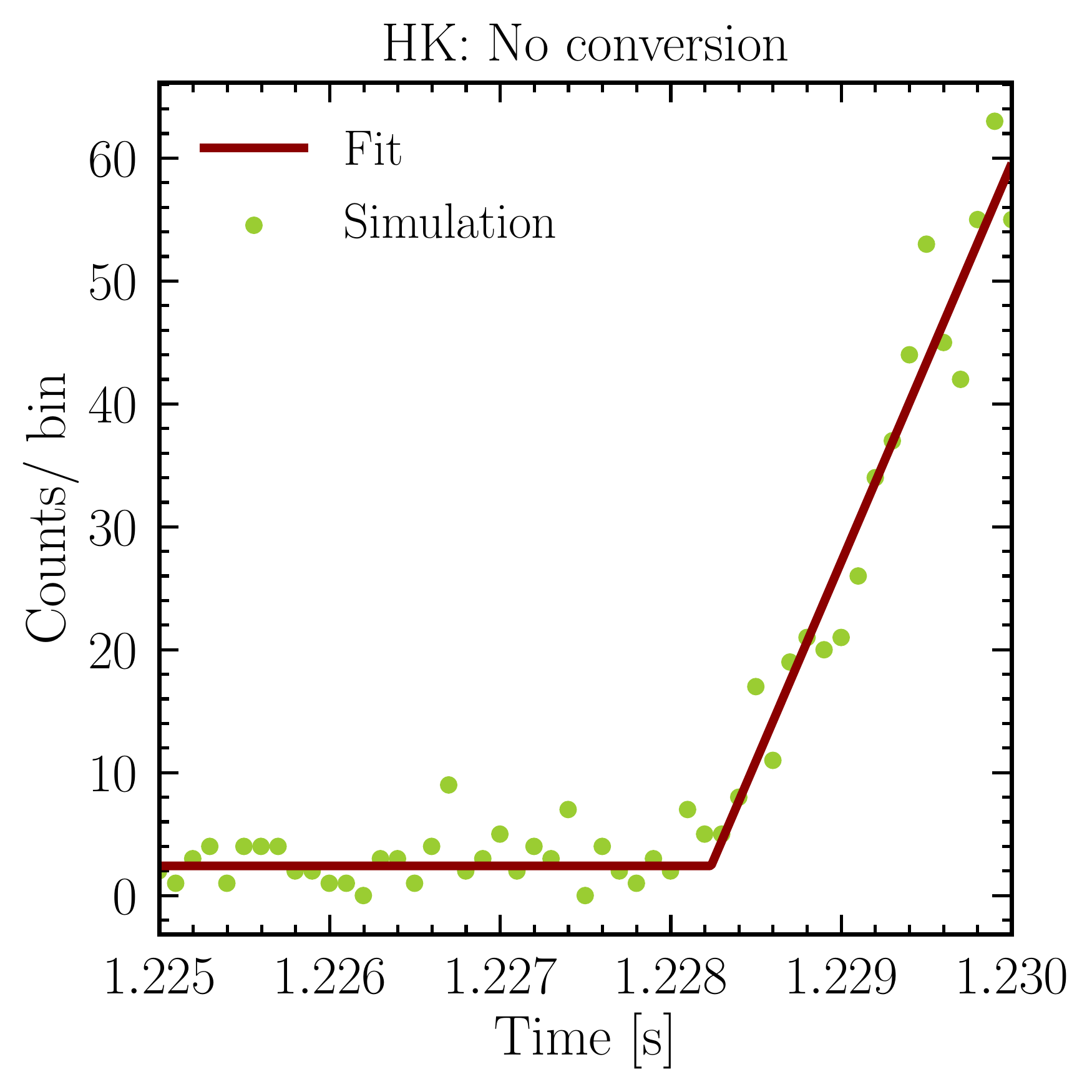}
	\includegraphics[width=0.9\columnwidth]{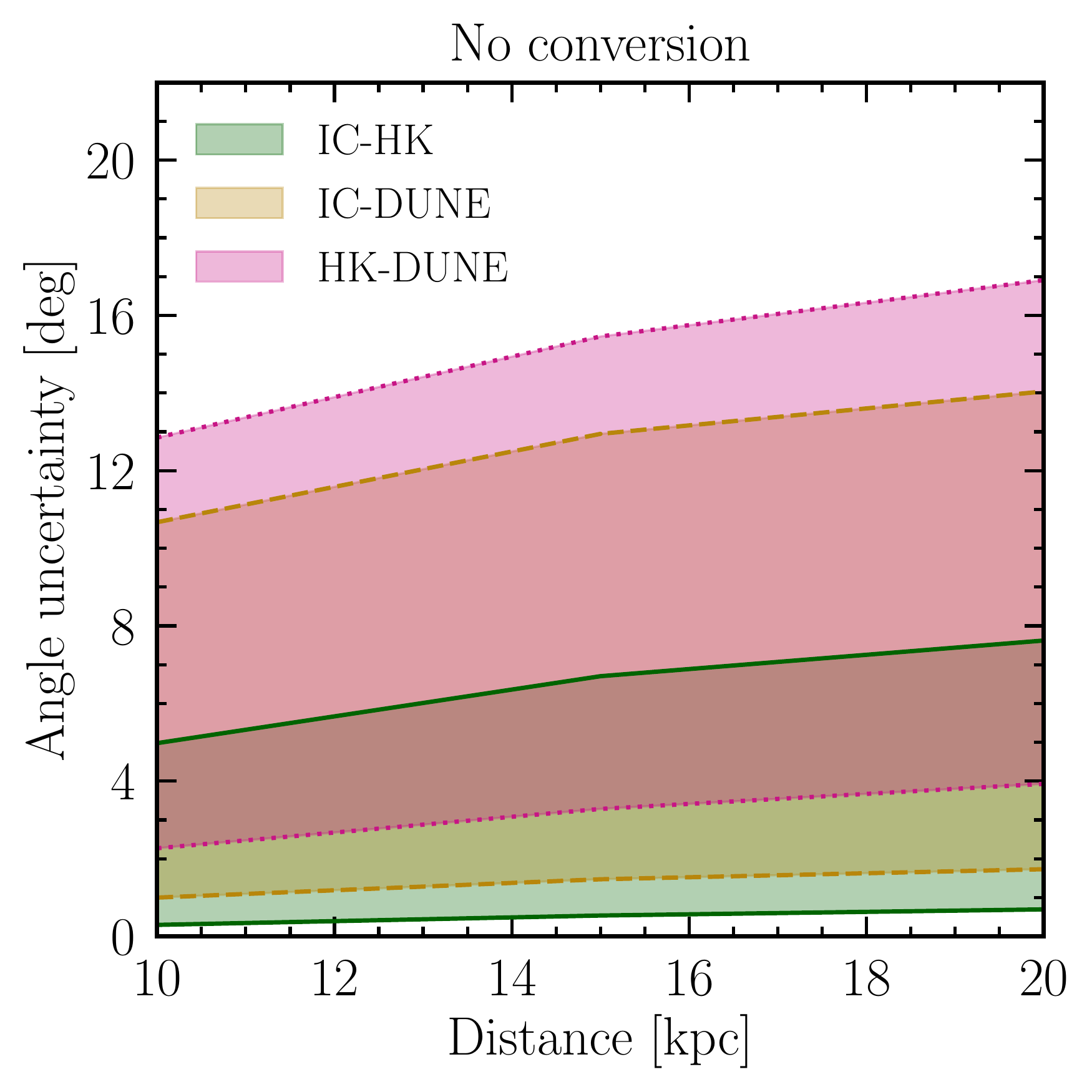}
	\caption{
	{\it{Upper panel:}} An example of the fitting procedure adopted to determine the rise time of the QHPT peak for a SN at 10 kpc, in HK detector. The green dots represent an instance of a randomly generated event rate binned with 0.1~ms time bin, while the solid brown line gives the best fit resulting from Eq.~\ref{eq:linear-fit}.
	{\it{Lower panel:}} Angle uncertainty range, computed in Eq.~\ref{eq:theta_error}, as a function of the distance in the no conversion scenario. Here we consider the angular error on the true SN location, by adopting the error on $\Delta t_{ij}^{\rm{true}}$ computed from Eq.~\ref{eq: delta_t_measured}. The minimum uncertainty for all three pairs of detectors is competitive with the resolution obtainable in water Cherenkov detectors by exploiting elastic scattering on electrons.
	}
	\label{Fig:Angle_uncertainty}
\end{figure}

The time delay recorded by different neutrino detectors can be used to infer the SN neutrino arrival direction with the triangulation method~\cite{Beacom:1998fj, Muhlbeier:2013gwa, Brdar:2018zds, Hansen:2019giq, Linzer:2019, Coleiro:2020vyj, Sarfati:2022}.
In order to use this method, we first calculate the time delays between pairs of detectors and their uncertainties by fitting the rise time of the QHPT peak in the neutrino event rates (see Fig.~\ref{fig:event-rate}).

Two factors limit the determination of the neutrino arrival time: first, the temporal resolution of the investigated detectors, and second, the theoretical uncertainty coming from the simulations' time step size. In our case, the latter is the limiting factor; thus, we bin the event rates in 0.1~ms bins for each detector, which is of the order of the temporal resolution of the simulation's data output in the vicinity of the neutrino peak caused by the QHPT.

Similarly, as in Ref.~\cite{Halzen:2009sm}, to calculate the rise time of the QHPT peak, we fit the neutrino event rates with the following linear function:
\begin{equation} 
\label{eq:linear-fit}
R_{\rm{exp}} = 
\begin{cases}
		    R_{\ast}, & \text{if $t < t_0$} \\
            R_{\ast}+a (t-t_0) , & \text{otherwise} \\
		 \end{cases}    \ ,
\end{equation}
where $R_{\ast}$ is the flat event rate before the QHPT peak appearance, $t_0$ is the start time of the QHPT peak that we are determining, and $a$ is a fitting parameter.
We apply the fitting function between $1.1$~s, where the flat signal dominates, and $1.23$~s, where we expect the QHPT burst to peak.
Note that the choice of a linear fitting function over, for example, an exponential one, is well motivated given the sharpness of the considered signal.
The upper panel of the Fig.~\ref{Fig:Angle_uncertainty} illustrates an example of the fitting procedure adopted for the HK event rate in case of no neutrino conversions.

By assuming Poisson statistics, for any combination among IC, HK, and DUNE, we generate $N_\mathrm{trials} = 3\times 10^{4}$ pairs of random burst signals. This number of trials $N_\mathrm{trials}$ ensures negligible variations in the obtained results. 
For each realization, we fit the rise time of the QHPT burst in each of the two detectors and calculate the time difference between them. Such time difference, which is due only to the fact that the considered detectors are not identical under one conversion scenario, is what represents the bias in Eq.~\ref{eq: delta_t_measured}  (see Sec.~\ref{sec:results-triangulation}).
The means and the standard deviations of the obtained in that way distributions are reported in Table~\ref{table:time_differences}, for the case of no conversion and full neutrino conversion. Histograms of the time distribution are also shown in Fig.~\ref{fig:time-distribiution}. The difference between the rise time and its error strongly depends on the conversion scenario in case of pairing with DUNE. The latter gives the largest uncertainty, mainly due to the low statistics. Nevertheless, the full conversion scenario is still significantly better than the no conversion case, both because of a larger number of events, and a more pronounced ratio peak to preburst signal.

\begin{table}[t]
	\caption{\label{table:time_differences} {\it Upper panel}: Time difference between pairs of neutrino detectors $-$IC, HK, and DUNE$-$ (corresponding to the bias $B_{ij}$ introduced in Eq.~\ref{eq: delta_t_measured}) in two flavor conversion scenarios considered in this work: no conversion and full conversion. The delay due to the time of flight between detectors located at different geographical locations is not taken into account (see Sec.~\ref{sec:results-triangulation}). The values reported here represent the mean and standard deviation of a set of $3\times 10^{4}$ trials. {\it Middle panel:} Minimum and maximum values of the angular uncertainty as defined in Eq.~\ref{eq:theta_error}, obtained for the true location, not considering the shift due to the bias. The impact of the bias on the localization of the SN is displayed in Fig.~\ref{Fig:Sky-maps-10kpc} and~\ref{Fig:Sky-maps-20kpc}. {\it Lower panel:} The median 95\% C.L. upper limit on the neutrino mass and its uncertainties calculated for $N_\mathrm{trials} = 5\times 10^4$ as in Eqs.~\ref{eq:lambda} and \ref{eq:uncertainty-mass}.}
	\begin{center}
	\begin{tabular}{ccc}
		\toprule
		{Detectors} & { No conversion} &{ Full conversion}\\
		\hline
		\multicolumn{3}{c}{$B_{ij}$ [ms]}\\
		\hline
		IC-HK & $-0.32\pm 0.10$ & $-0.32 \pm 0.10 $\\
		IC-DUNE & $-0.11\pm 0.48$ & $-0.27 \pm 0.20$\\
		HK-DUNE & $0.22\pm 0.50$ & $0.05 \pm 0.22$\\
		\hline
		\multicolumn{3}{c}{$\delta(\theta_{ij})$ (min, max) [deg]}\\
		\hline
		IC-HK & $(0.30,5.00)$ & $(0.29,4.90)$\\
		IC-DUNE & $(1.00,10.67)$ & $(0.41,6.90)$\\
		HK-DUNE & $(2.27,12.85)$ & $(1.00,8.54)$\\
		\hline
		\multicolumn{3}{c}{95\% C.L. upper limit on $m_\nu$ [eV]}\\
		\hline
		{IC} & {$0.16_{-0.04}^{+0.03}$} & {$0.21_{-0.05}^{+0.05}$}\\
		{HK} & {$0.22_{-0.06}^{+0.05}$} & {$0.30_{-0.09}^{+0.07}$}\\
		{DUNE} & {$0.80_{-0.29}^{+0.21}$} & {$0.58_{-0.19}^{+ 0.14}$}\\
		\toprule
	\end{tabular}
	\end{center}
\end{table}

\subsection{The triangulation method}
\label{sec:results-triangulation}

\begin{figure*}[t]
\includegraphics[width=1\columnwidth]{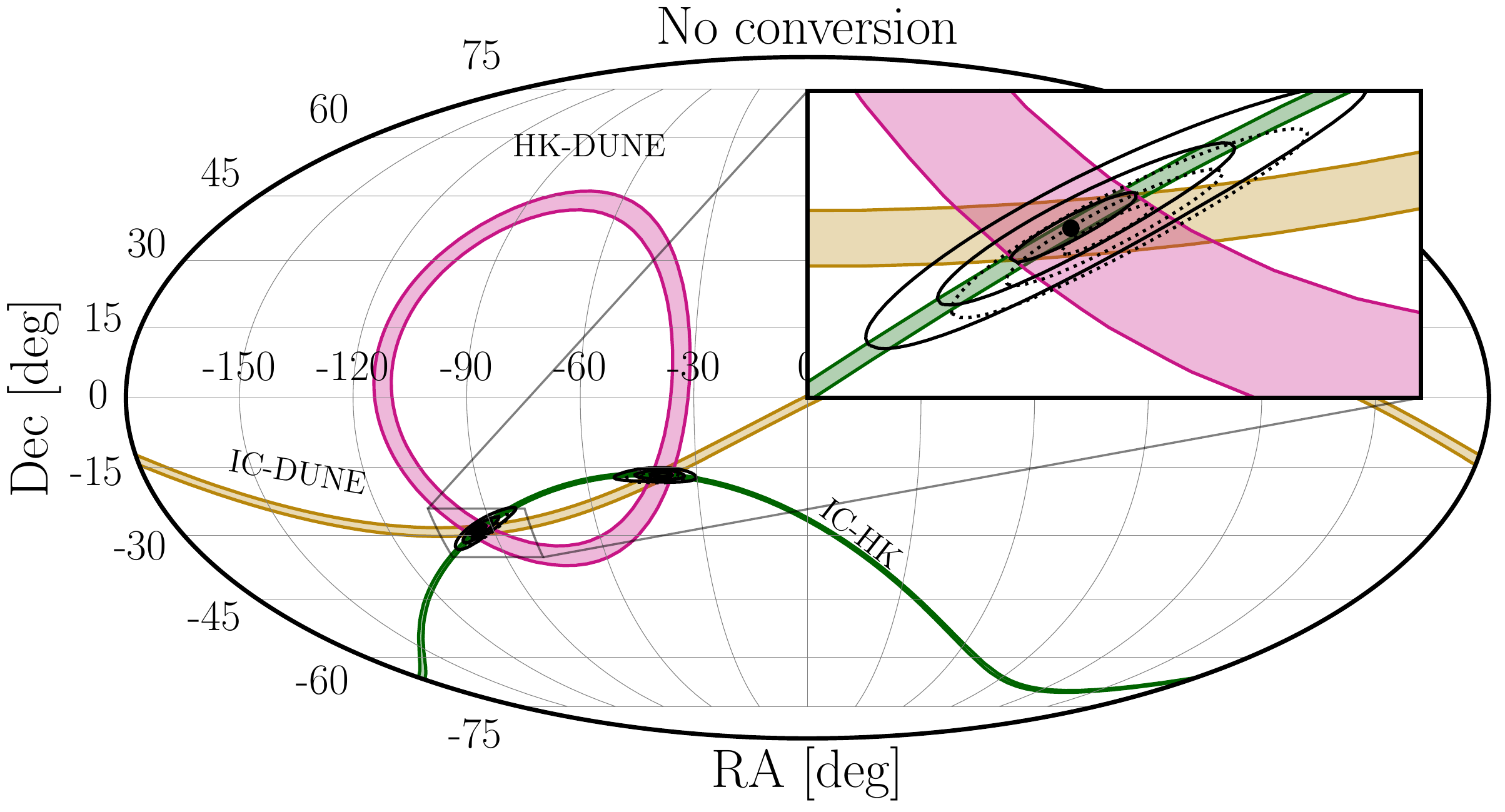} 
\includegraphics[width=1\columnwidth]{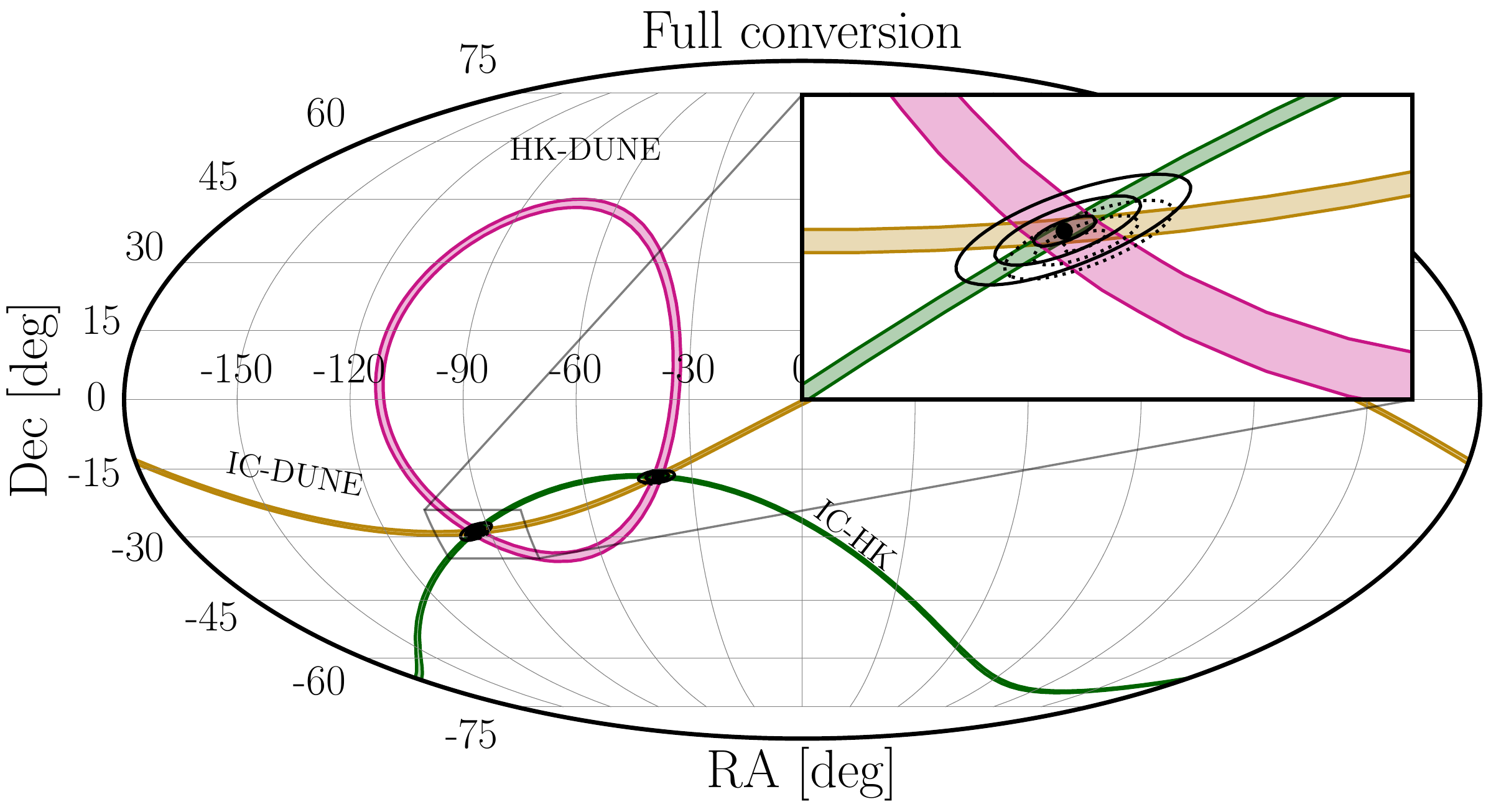}

\caption{The colored regions represent the 1$\sigma$ areas on the sky constraining the location of the QHPT SN at a distance 10 kpc and true location at the galactic center. The black solid (dotted) lines in the inset show the 1, 3, and 5$\sigma$ confidence regions for the SN localization obtained by using timing information from IC, HK and DUNE without (with) the bias correction described in Sec.~\ref{sec:results-triangulation}. The existence of the bias leads to a significant weakening of the pointing accuracy, while the full conversion represents the scenario with the best angular precision obtainable for QHPT peak.}
\label{Fig:Sky-maps-10kpc}
\end{figure*} 

We have established the bias in the measurement of the rise time due the different detector properties for the set conversion scenario. Now, in order to employ the triangulation method, we compute the temporal delay between the arrival times of the neutrino pulse in different telescopes. For two detectors on Earth located at $\textbf{r}_i $ and $\textbf{r}_j$, the recorded time delay can be expressed as:
\begin{equation}
\label{eq:time-difference}
\Delta t^{\rm{true}}_{ij} = \frac{(\textbf{r}_i - \textbf{r}_j) \cdot \textbf{n}}{c} = \frac{{\rm{D}}_{ij}\cos\theta}{c}\quad ,
\end{equation}
where $c$ is the speed of light, ${\rm{D}}_{ij}=|\textbf{r}_i - \textbf{r}_j|$ the distance between the detectors listed in Table~\ref{table:detectors}, and $\textbf{n}$ is the unit vector that indicates the direction from which the neutrinos arrive. For a SN occurring at right ascension $\alpha$ and declination $\delta$, this vector is given by:
\begin{equation}
	\textbf{n} = (- \cos \alpha \cos \delta, - \sin \alpha \cos \delta, -\sin \delta)
\end{equation}
We assume that the SN is located at the galactic center with $\alpha=-94.40$~deg and $\delta=-28.92$~deg, and the collapse happens on the vernal equinox at noon coordinate universal time (UTC).
Following Refs.~\cite{Linzer:2019, Sarfati:2022} we define the total time delay as:
\begin{equation}
\label{eq: delta_t_measured}
	\Delta t^{\rm{measured}}_{ij}=\Delta t^{\rm{true}}_{ij}+B_{ij}
\end{equation}
where $B_{ij}$ is the bias provided in Table~\ref{table:time_differences}. The uncertainty on $\Delta t^{\rm{true}}_{ij}$ is obtained by adding in quadrature the one on $B_{ij}$ and $\Delta t^{\rm{measured}}_{ij}$, resulting in a factor $\sqrt{2}$ increase with respect to $\Delta t^{\rm{measured}}_{ij}$. 

Given a time delay $\Delta t_{ij}$ with an error $\delta(\Delta t_{ij})$, the uncertainty on the determination of $\theta$ (Eq.~\ref{eq:time-difference}) is
\begin{equation}
	\delta (\cos\theta_{ij})=\frac{\delta (\Delta t_{ij}) \, c}{{\rm{D}}_{ij}}
\end{equation}
 which can be estimated as follows~\cite{Beacom:1998fj, Linzer:2019}:
\begin{equation} 
	\label{eq:theta_error}
	\delta(\theta_{ij}) \approx 
	\begin{cases}
		\delta(\cos\theta_{ij})/\sin\theta_{ij} & \text{if $\sin\theta_{ij} >\sqrt{\delta(\cos\theta_{ij})}$} \\
		\sqrt{2\delta(\cos\theta_{ij})} , & \text{for}\,\theta_{ij}\ll \delta(\cos\theta_{ij}) \\
	\end{cases}    \ .
\end{equation}
These are the two extremal angular uncertainties, the former one for large angles (up to $\theta\sim 90^{\circ}$), and the latter for small ones ( $\theta\sim 0^{\circ}$), which limit the determination of the SN location. Their values are reported in Table~\ref{table:time_differences} for each pair of detectors for SN at a distance of~10~kpc.
As opposed to no conversion, in the case of full conversion the pairing with DUNE improves significantly the pointing precision because of a smaller error on $\Delta t^{\rm{true}}_{ij}$. All detector pairs provide a competitive minimum angular uncertainty if compared to using elastic scattering on electrons.
The bottom panel of the Fig.~\ref{Fig:Angle_uncertainty} shows the range of angular uncertainties as a function of the distance to the SN for the case of no conversion. Due to a decrease in the number of events in all considered detectors with increasing distance to the SN the error on the determination of the SN location grows.

\subsection{Combined analysis}
\label{sec:combined-localization}

In order to calculate the confidence regions of the SN localization in the sky, we perform a combined analysis using the results from all three detectors. Similarly as in Ref.~\cite{Sarfati:2022}, to take into account correlations among the detectors, we construct the following $\chi^{2}$ function~\cite{Barlow}:
\begin{equation}
	\label{eq: chis_squared}
	\chi^{2}(\alpha,\delta)=(\Delta\textbf{t}-\Delta\textbf{t}^{\rm{true}}(\alpha,\delta))^{T}C^{-1}_{\rm{true}}(\Delta\textbf{t}-\Delta\textbf{t}^{\rm{true}}(\alpha,\delta))
\end{equation}
where $\Delta\textbf{t}^{\rm{true}}(\alpha,\delta)$ is the theoretically expected time difference for a SN located at $(\alpha,\delta)$, ${\Delta\textbf{t}=\Delta\textbf{t}^{\rm{true}}(\alpha^{\prime},\delta^{\prime})+\textbf{B}}$ is the expected measured time difference for the true SN located at $(\alpha^{\prime},\delta^{\prime})$ as defined in Eq.~\ref{eq:time-difference}, and $C_{\rm{true}}$ the covariance matrix. The minimum of the $\chi^{2}$ function gives the best estimate for the angles $(\alpha,\delta)$ of the real SN location in the sky. The resulting sky regions for each pair of detectors are shown in Fig.~\ref{Fig:Sky-maps-10kpc} at $1\sigma$, $3\sigma$, and $5\sigma$ C.L., for both oscillation scenarios.

The unique determination of the SN direction on the sky requires the information from at least four detectors. In our work, to demonstrate the potential of using the QHPT peak to determine the SN direction we use the information from three neutrino detectors, which limits the SN localization to two solutions (see Figs.~\ref{Fig:Sky-maps-10kpc} and \ref{Fig:Sky-maps-20kpc}). Addition of the fourth detector with similar characteristics as the ones employed in our works, e.g., Super-Kamiokande (SK)~\cite{Super-Kamiokande:2016kji} or Jiangmen Underground Neutrino Observatory (JUNO)~\cite{JUNO:2022hxd} will allow to determine a unique location of the SN.

\subsection{Sensitivity to the absolute neutrino mass}
\label{sec:results-neutrino-mass}

The presence of any sharp feature in the SN neutrino signal may be used to constrain the absolute neutrino mass~\cite{Zatsepin:1968kt, Beacom:2000ng, Loredo:2001rx, Nardi:2003pr, Nardi:2004zg, Pagliaroli:2010ik, Lu:2014zma, Rossi-Torres:2015rla, Hansen:2019giq, Pompa:2022cxc} because massive neutrinos introduce distinct energy-dependent time delays (SNe emit non-monoenergetic neutrino spectrum, see Eq.~\ref{eq:energy-distribiution}).
The neutrino flux (Eq.~\ref{eq:flux}) arriving at the Earth is modified by the neutrino mass as $F (E_{\nu}, t)\rightarrow F (E_{\nu}, t^\prime; m_\nu)$, with $t^\prime = t - \Delta t (m_{\nu})$, where $\Delta t$ is energy dependent time delay introduced by nonzero $m_\nu$ given by Eq.~\ref{eq:time-delay}. The nonzero neutrino mass causes low-energy neutrinos to arrive later than high-energy ones. Therefore, the sharper the expected feature, the better the mass limit because more massive neutrinos smear the peak out onwards later times. 

\begin{figure*}[t]
\includegraphics[width=\columnwidth]{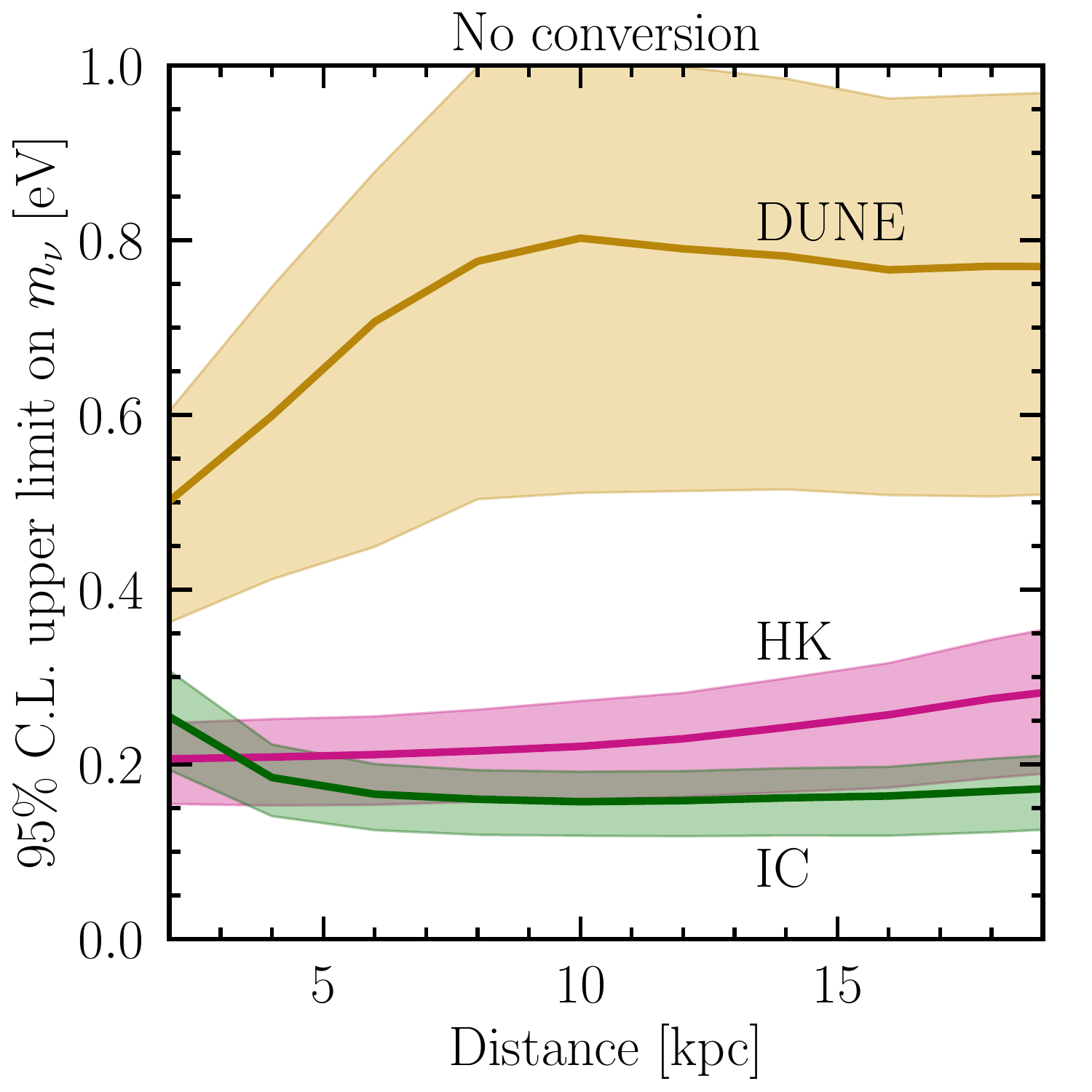}
\includegraphics[width=\columnwidth]{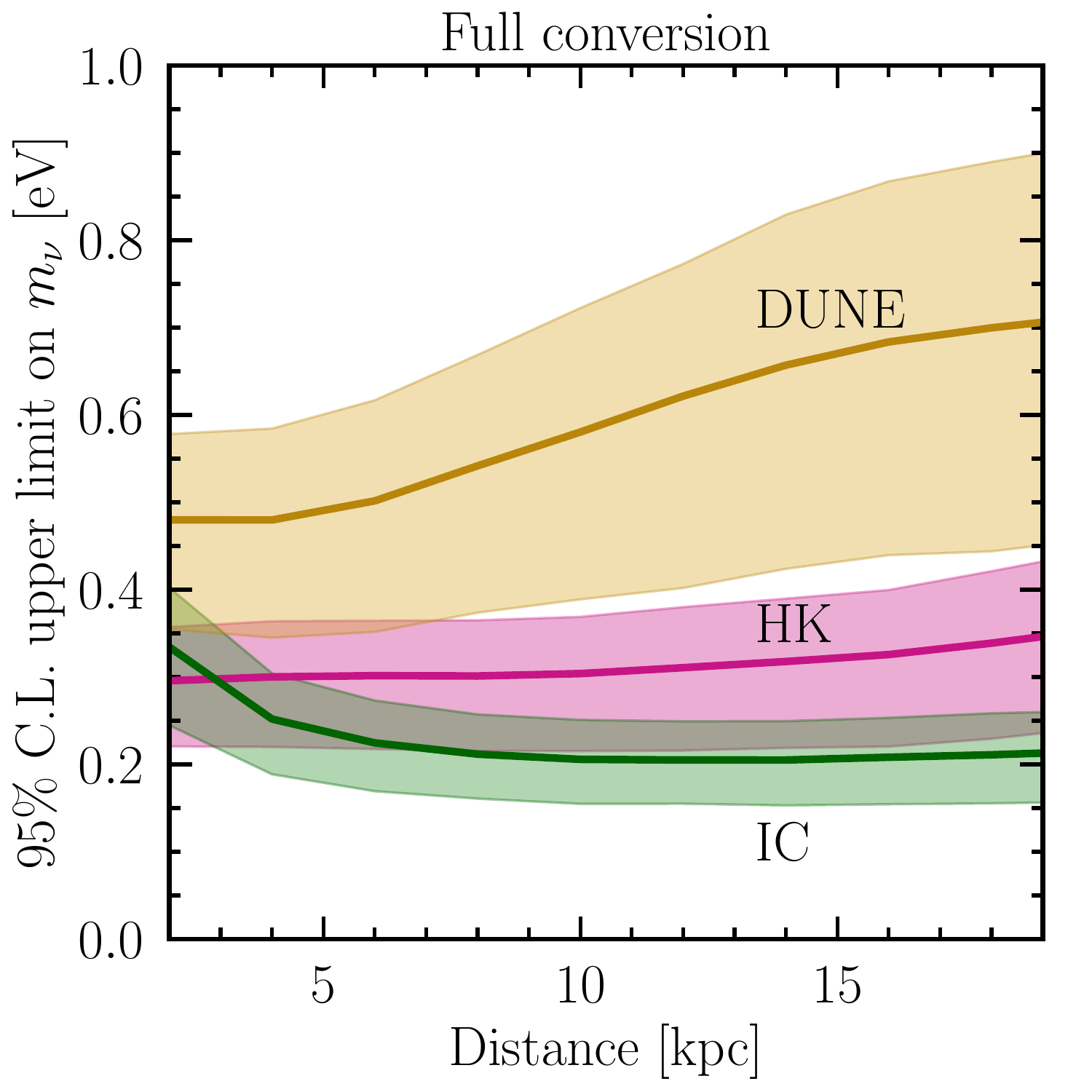}
\caption{The median expected 95\% C.L. upper limit on neutrino mass from QHPT SN neutrino peak for HK (pink), IC (green), and DUNE (yellow) in the no conversion (left panel) and full conversion (right panel), oscillation scenarios. The shaded bands represent the upper and lower $1\sigma$ uncertainties on the calculated 95\% C.L. limits. Due to its smaller volume and lower efficiency DUNE sets the weakest limits among the considered detectors.}
\label{Fig:neutrino_mass_limit}
\end{figure*}

To find the 95\%~C.L. upper limits on the neutrino mass and their uncertainties, we use the likelihood ratio method and assume that the calculated ratios can be well approximated by the $\chi^2$ distribution with 1 degree of freedom under the Wilks' theorem~\cite{Wilks:1938dza} (see Appendix~\ref{app:cross-cheks} for the results without that assumption). 
The sensitivity to the neutrino mass is thus calculated using the following formula
\begin{equation}
\label{eq:Wilks-theorem}
\Delta \chi^2 \approx -2 \log(\Lambda)   \ 
\end{equation}
with the likelihood ratio given by~\cite{Cowan:2010js}
\begin{equation}
\label{eq:lambda}
    \Lambda = \begin{cases} \frac{L(m_{\nu}^\mathrm{best})}{L(m_{\nu})}, & \mathrm{if~} m_{\nu}^{\mathrm{best}} \leq m_{\nu} \\
    1, & \mathrm{if~} m_{\nu}^{\mathrm{best}} > m_{\nu} \ ,
    \end{cases}
\end{equation}
where $m_{\nu}^{\mathrm{best}}$ is the best fit value corresponding to ${\mathrm{min}}\{L(m_\nu)\} = L(m_\nu^\mathrm{best})$.
The likelihood functions $L$ is expressed by the Poisson distributions as
\begin{equation}
\label{eq:likelihood}
    L(m_\nu) = \prod_{i=0}^{N} P(\lambda = N_i(m_\nu); k = N_i^\mathrm{MC})
\end{equation}
where $i$ is the bin number, $m_\nu = \{ m_{\nu}, m_{\nu}^{\mathrm{best}} \}$, $N_i (m_\nu)$ is the number of expected events for $m_\nu$ hypothesis, and $N_i^\mathrm{MC}$ is the simulated number of observed events drawn for the zero neutrino mass hypothesis.

Fixed a neutrino detector, a conversion scenario, and a distance to the SN, we draw the number of observed events from the $m_\nu=0$ hypothesis in the time window 1.225 - 1.233 s with 0.1~ms bins, and use Eqs.~\ref{eq:lambda}--\ref{eq:likelihood} to calculate the $\Delta\chi^2$ for $m_\nu = [0, 3~\mathrm{eV}]$. The neutrino mass, which corresponds to the critical value of the $\chi^2$ distribution for the 95\%~C.L. (equal to 3.81 for 1 degree of freedom (d.o.f)), is the calculated upper limit from a single trial.  
We repeat this procedure many times ($N_\mathrm{trials} = 5\times 10^4$) to construct the probability density distribution of the limits from single trials (Fig.~\ref{Fig:mass_distribiution} in Appendix~\ref{app:histograms-mass-limit}). The median of the obtained distribution is the final 95\%~C.L. on the absolute neutrino mass.
Since the distributions are not Gaussian (see Fig.~\ref{Fig:mass_distribiution} in Appendix~\ref{app:histograms-mass-limit}), we determine the 68\% C.L. uncertainty on the median 95\% C.L. upper limit via the Feldman-Cousins procedure to ensure consistent coverage~\cite{Feldman:1997qc}.
The lower ${m_\nu^\mathrm{low}}$ and upper ${m_\nu^\mathrm{up}}$ uncertainties on the 95\% C.L. upper limit are the solutions to
\begin{equation}
\label{eq:uncertainty-mass}
\int_{m_\nu^\mathrm{low}}^{m_\nu^\mathrm{up}} \rho (m_\nu) \; dm_\nu = 0.682 \ ,
\end{equation}
where $\rho (m_\nu)$ is the constructed probability density distribution and $\rho ({m_\nu^\mathrm{up}}) = \rho ({m_\nu^\mathrm{low}})$. The calculated 95\% C.L. upper limits for a SN at a distance 10~kpc for all three considered detectors, and both oscillation scenarios are summarized in Table~\ref{table:time_differences}. 

Figure~\ref{Fig:neutrino_mass_limit} shows the 95\%C.L. upper limits on the neutrino mass and their uncertainty from the HK (pink), IC (green), and DUNE (yellow) as a function of the distance to the SN for both considered neutrino conversion cases.
These limits and their uncertainty range were calculated for $N_\mathrm{trials}=10^4$  to save the computational expenses. We have checked that increasing the number of trials does not affect the limits and their uncertainties more than approximately 2\%. For example, in the the full conversion scenario from HK detector for $N_\mathrm{trials}=10^4$ and SN at a distance 10~kpc the median, mean, lower uncertainty, and upper uncertainty of the limit are 0.304, 0.312, 0.215, and 0.369, respectively (see the left panel of Fig.~\ref{fig:cross-checks} in Appendix~\ref{app:cross-cheks}), whereas for $N_\mathrm{trials}=5\times10^4$ these numbers change to 0.305, 0.313, 0.215, and 0.373 (see the Table~\ref{table:time_differences} and the lower right panel of the Fig.~\ref{Fig:mass_distribiution} in Appendix~\ref{app:histograms-mass-limit}).

The IC experiment results in the best limits on the neutrino mass from QHPT peak for SN at distances greater than approximately 5~kpc due to its largest volume and lowest energy threshold among the considered detectors. For SN occurring closer to the Earth, however, the HK detector yields better limits because the dead-time efficiency significantly reduces the neutrino signal in the IC during the QHPT peak. DUNE, owing to the smallest volume and lowest effective threshold, is expected to provide the weakest limits on the neutrino mass. 

In addition, in the case of no conversion, the calculated limits are stronger in HK and IC than in the full conversion case, and the opposite is true for DUNE. This is due to the hierarchy of the luminosities and mean energies of the different neutrino flavors emitted from the SN (see Fig.~\ref{Fig:luminosity_mean_energy}). The mean energies and luminosities for $\nu_e$ are the lowest during the QHPT peak; therefore, the limits from DUNE in the no conversion case are worse than in full conversion.
On the other hand, the mean energies of $\bar\nu_x$ are larger than $\bar\nu_e$, but the luminosity of $\bar\nu_e$ is larger than $\bar\nu_x$. Combining that with shorter time delays for more energetic neutrinos (Eq.~\ref{eq:time-delay}) leads to better limits in the no conversion than full conversion case for IC and HK. The effect is, however, not as striking as for DUNE.  

We have also checked that including a normalization uncertainty in the likelihood calculation and marginalizing over it does not affect our results significantly. This is because the nonzero neutrino masses reduce the left side of the QHPT neutrino peak while extending the right slope, effectively widening and flattening the peak, an effect that cannot be mimicked by changing the normalization.

\section{Discussion and Conclusions}
\label{sec:Conclusions}

Core-collapse supernovae are one of the most complex phenomena in the universe. Not only are they a site of production of the heavy elements which enable the existence of life, but their cores are also one of the densest environments we can indirectly probe. At such densities, the matter may no longer consist only of hadronic degrees of freedom but undergo a phase transition to quark matter. It has been shown that such transition can cause a prominent burst of $\bar{\nu}_{e}$ in the early postbounce phase after the neutronization burst release. The detection of such a sharp feature in the neutrino signal from a core-collapse supernova at the Earth would not only strongly back the existence of the quark-hadron phase transition, but also allow us to set stringent limits on the neutrino masses and help to identify the supernova location by the triangulation technique.

In this work, we show that triangulating supernova using the quark-hadron phase transition $\bar{\nu}_{e}$ peak can significantly improve the pointing precision of the supernova localization within the galactic neighborhood compared to using the neutronization peak. Given the characteristic flavor composition, energy dependence of the sharp $\bar{\nu}_{e}$ burst, and considered detectors, the case with full conversion between the neutrino flavors leads to the best determination of the collapsing star's location, i.e., $\sim0.3^{\circ} - 9.0^{\circ}$ angular uncertainty for a supernova at distance of 10~kpc. In the case of no conversion between the flavors the pointing precising reduces to $\sim 0.3^{\circ}-13.0^{\circ}$. 

Our method leads up to $\sim 4.5-10$ times improvement in comparison to utilizing the neutronization burst~\cite{Beacom:1998fj, Muhlbeier:2013gwa, Brdar:2018zds, Hansen:2019giq, Linzer:2019, Coleiro:2020vyj} to triangulate supernova, while comparable results have been found by using the endpoint of the neutrino signal for progenitors directly forming black holes~\cite{Muhlbeier:2013gwa, Brdar:2018zds, Hansen:2019giq, Sarfati:2022}.
Even though better techniques to pinpoint the SN location exist, i.e., using the directionality of interaction channels such as neutrino elastic scattering off electrons, an optimal approach is to gather all the available information by exploiting all existing methods and complementary strategies.

We also show that observation of the sharp $\bar{\nu}_{e}$ burst associated with the quark-hadron phase transition can set competitive limits on the neutrino mass. We find that IC can reach the 95\% C.L. sensitivity of 0.16~eV, HK 0.22~eV, and DUNE 0.58~eV for supernova at a distance 10 kpc from the Earth for favorable conversion scenarios, i.e., no conversions for HK and IC, and full flavor swap for DUNE. The limits for the opposite conversion scenarios are 0.21~eV, 0.30~eV, and 0.80~eV for IC, HK, and DUNE, respectively.
The calculated upper limits on the neutrino mass are more stringent than the ones from the laboratory experiments 0.8~eV~\cite{KATRIN:2021uub} and the limits~$0.5-1.5~\mathrm{eV}$ obtained from the supernova neutronization burst analyses~\cite{Nardi:2003pr, Nardi:2004zg, Pagliaroli:2010ik, Lu:2014zma, Rossi-Torres:2015rla, Hansen:2019giq, Pompa:2022cxc}. In addition, they are comparable to the bounds found using the end time of the neutrino signal from a black-hole-forming core-collapse supernova in cases with low accretion~\cite{Beacom:2000ng, Hansen:2019giq}. We expect that detectors with comparable characteristics that encompass energy thresholds, time resolutions and effective volumes such as, e.g., SK or JUNO, can yield similar results to the ones found in our paper.

The existence of the hadron-quark phase transition in the core-collapse supernova may not only be imprinted in the neutrino signal, but also gravitational waves~\cite{Blacker:2020nlq}. In such a case, the time difference between the gravitational wave and neutrino signals may also be used to ascertain the neutrino mass limits, as in the neutronization burst and BH formation cases~\cite{Pagliaroli:2009qy, Hansen:2019giq}. 
Moreover, the quark-hadron phase transition can result in increased entropy and neutron richness in the neutrino-driven wind~\cite{Fischer:2020xjl}, setting an optimal condition for the $r-$process.

In our work, we rely on a specific model of the core-collapse supernova with a $\bar\nu_e$ burst released by the quark-hadron phase transition. The results presented here, however, should not change significantly for different models because the heights and widths of the peak are comparable in those, see e.g., Refs.~\cite{Dasgupta:2009yj, Jakobus:2022ucs}. 
Nevertheless, it would be interesting to see a study focused on how the theoretical uncertainties inherent in the modeling of the phase transition in the supernova affect the emitted neutrino spectra. 

\begin{acknowledgments}

We are grateful for helpful discussions with George Fuller, Massimiliano Lincetto, Shashank Shalgar, Toshio Suzuki, Mark Vagins, Anna Watts, and especially Peter Denton, Rasmus Hansen, and Kate Scholberg.
We also thank Tobias Fischer for useful insights and access to the supernova-models.
T.P would like to thank Irene Tamborra for the support during the course of this project. 
This work was supported in part by the National Science Foundation Grants No. PHY-2020275 and PHY-2108339. T.P acknowledges support from the Carlsberg Foundation (CF18-0183). A.M.S. and A.B.B would like to thank Kavli Institute for Theoretical Physics (KITP) for the hospitality during this work. KITP issupported in part by the National Science Foundation under Grant No. NSF PHY-1748958.
We acknowledge the use of the matplotlib~\cite{Hunter:2007ouj} and Astropy python packages~\cite{astropy:2013, astropy:2018}.
\end{acknowledgments}

\appendix

\section{Histograms: timing the neutrino signal}
\label{app:histograms-triangulation}

\begin{figure*}[t]
\includegraphics[width=2\columnwidth]{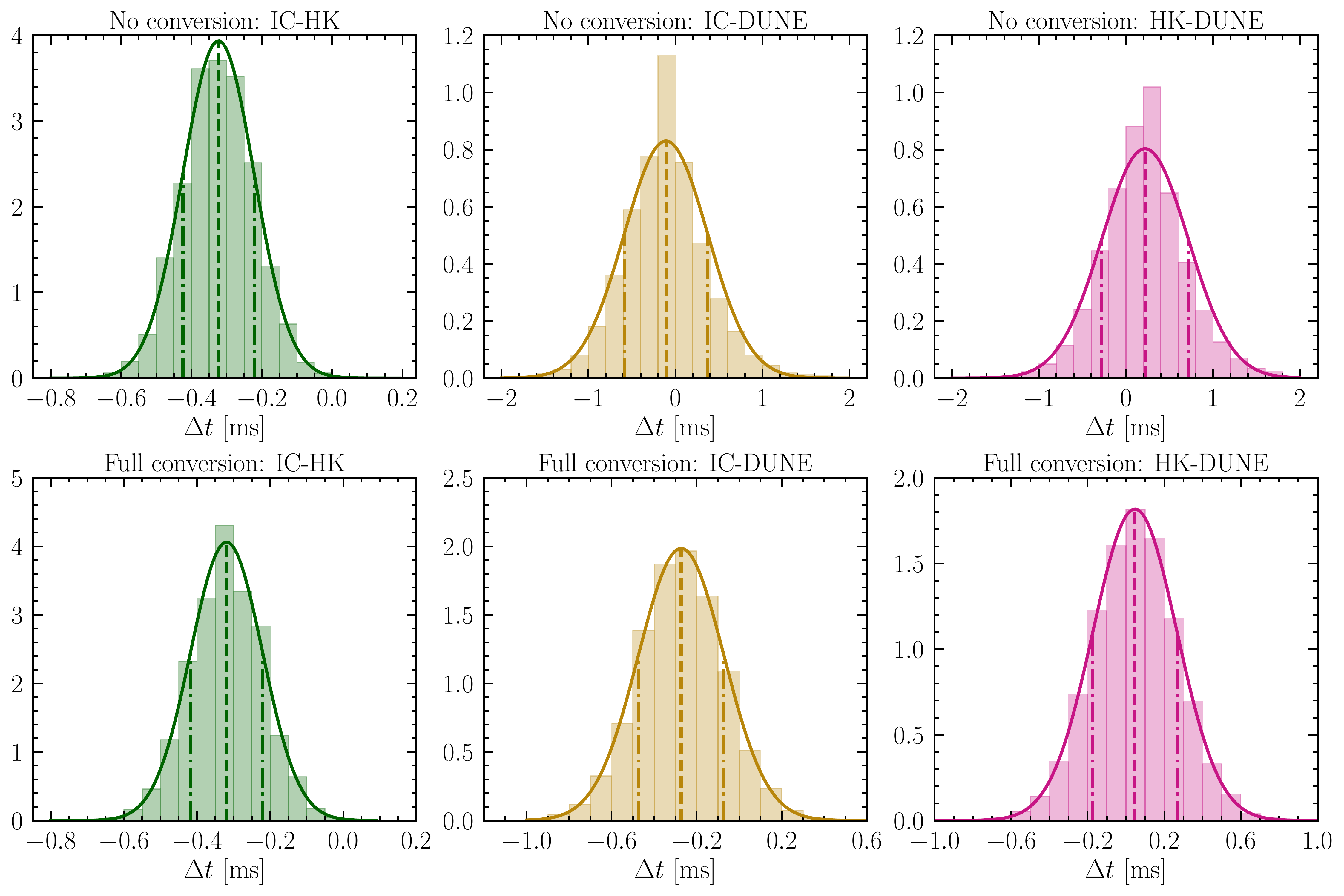}

\caption{The distribution of bias for the pairs of the detectors IC-HK (green), IC-DUNE (yellow), and HK-DUNE (pink) in the "No conversion" case in the upper and "Full conversion" lower panels.}
\label{fig:time-distribiution}
\end{figure*} 

We show the constructed distribution of the biases between the pairs of the considered detectors for the QHPT neutrino burst time determination in Fig.~\ref{fig:time-distribiution} for a SN at a distance of 10~kpc. In all of the displayed cases, the calculated distribution are well fitted with the Gaussians (solid lines) characterized by the mean (dashed lines) and standard deviation (dash-dotted lines) from $N_\mathrm{trial} = 3\times10^4$.

\begin{figure*}[t]
\includegraphics[width=1\columnwidth]{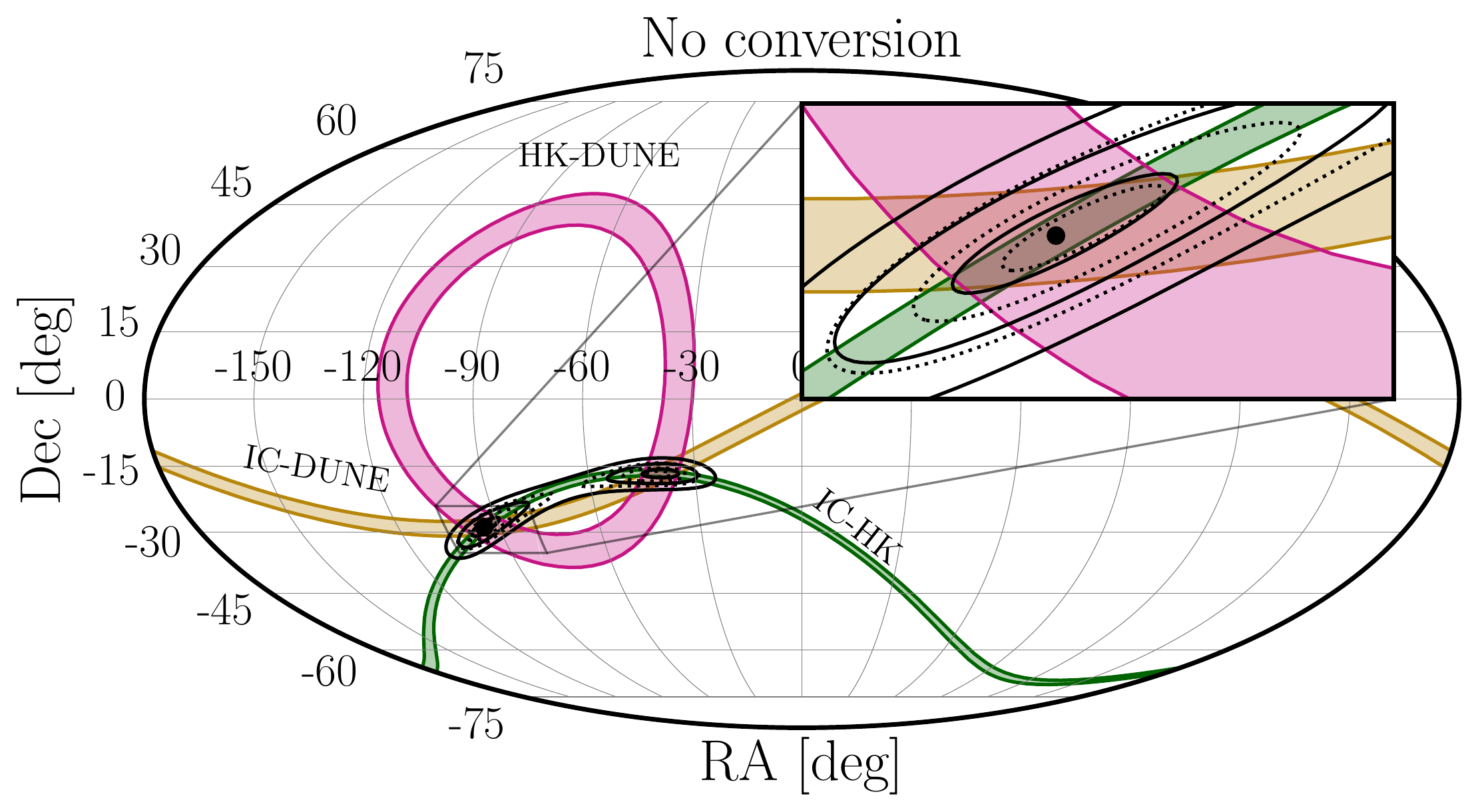} 
\includegraphics[width=1\columnwidth]{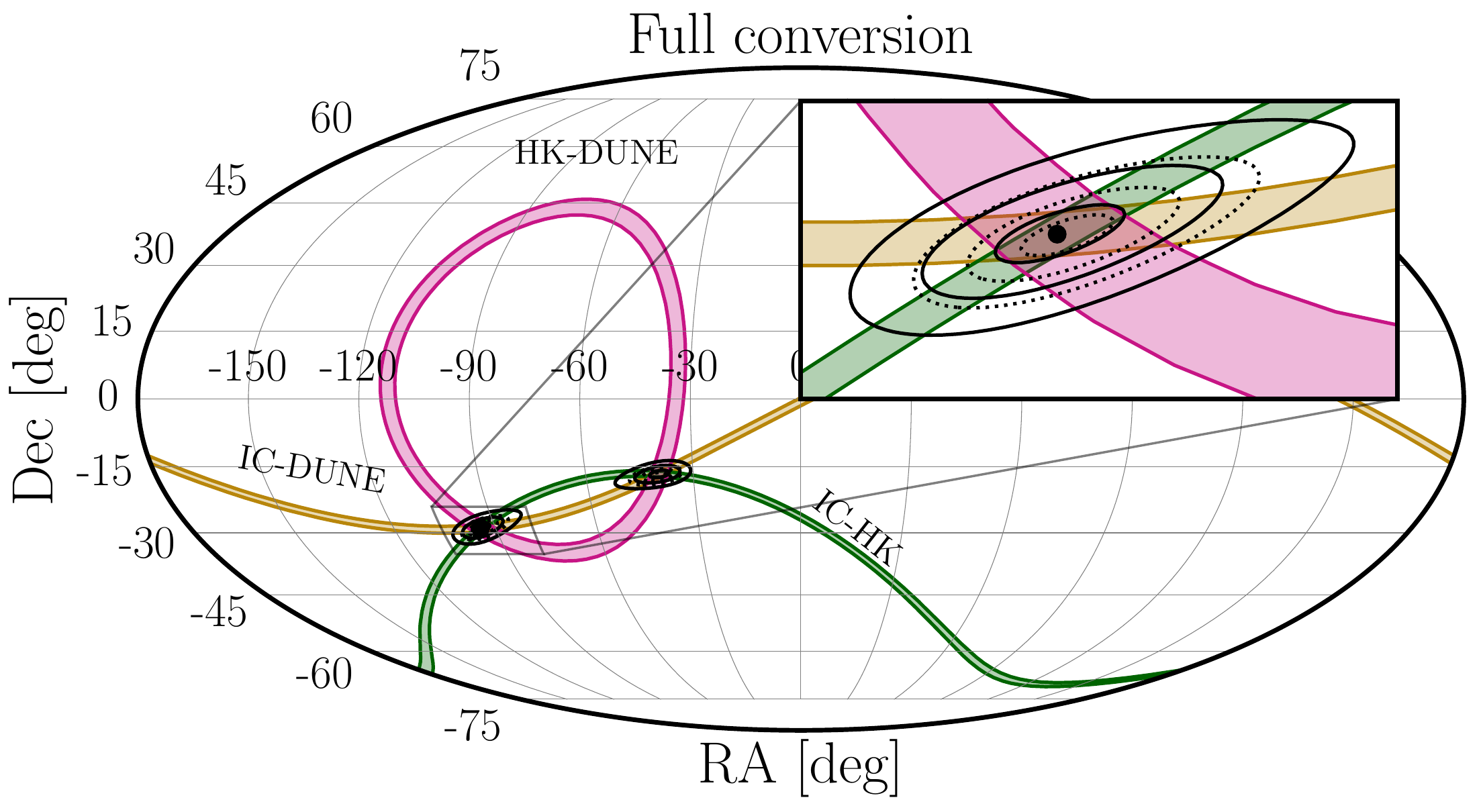} 
\caption{Same as in Fig.~\ref{Fig:Sky-maps-10kpc}, but for a distance of 20 kpc. The pointing precision reduces with an increasing distance to the SN.}
\label{Fig:Sky-maps-20kpc}
\end{figure*} 

Additionally, in Fig.~\ref{Fig:Sky-maps-20kpc} we show how the determination of the SN localization on the sky weakens in case the distance to the Earth increases to 20~kpc.

\section{Histograms: neutrino mass limit}
\label{app:histograms-mass-limit}
\begin{figure*}[t]
\includegraphics[width=0.65\columnwidth]{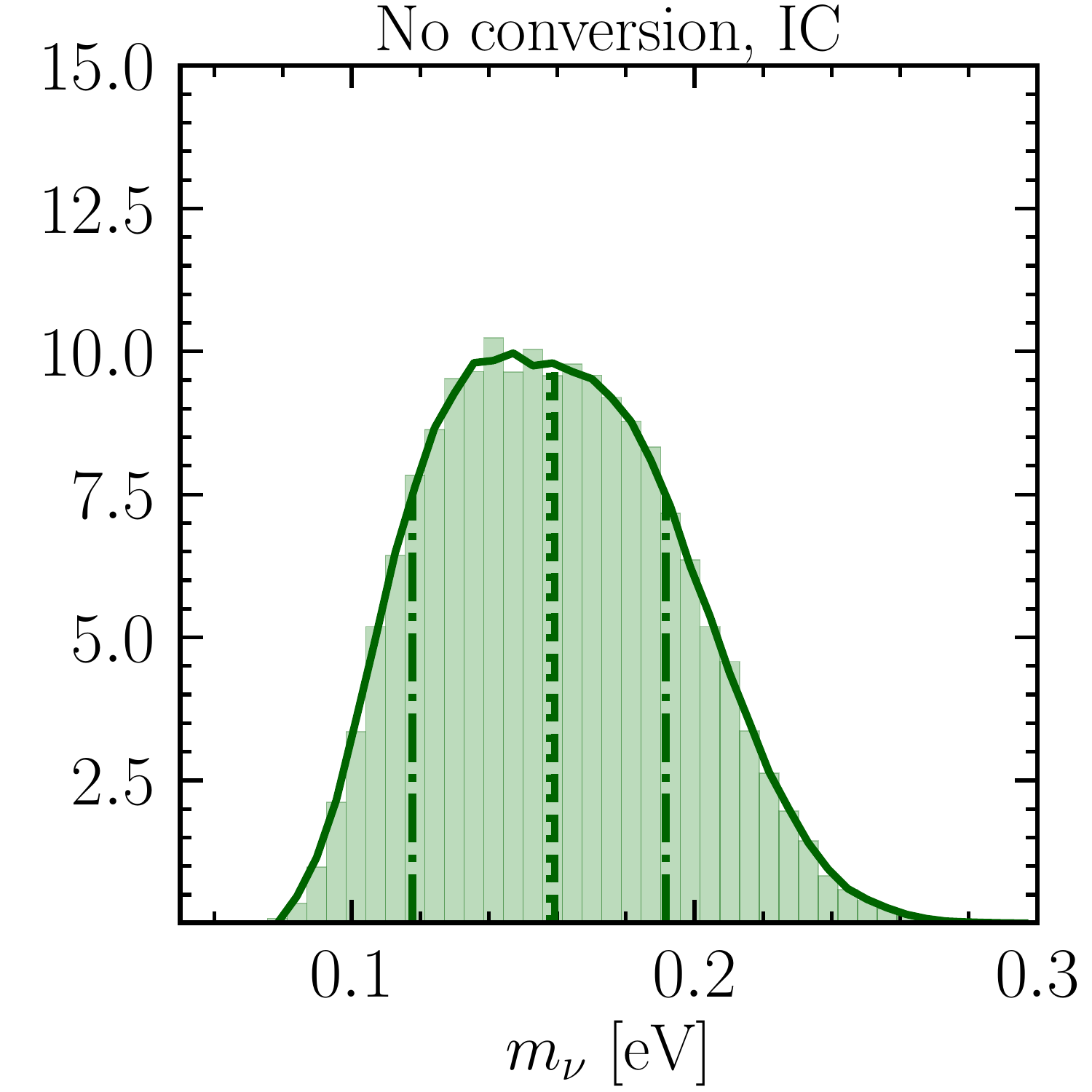}
\includegraphics[width=0.65\columnwidth]{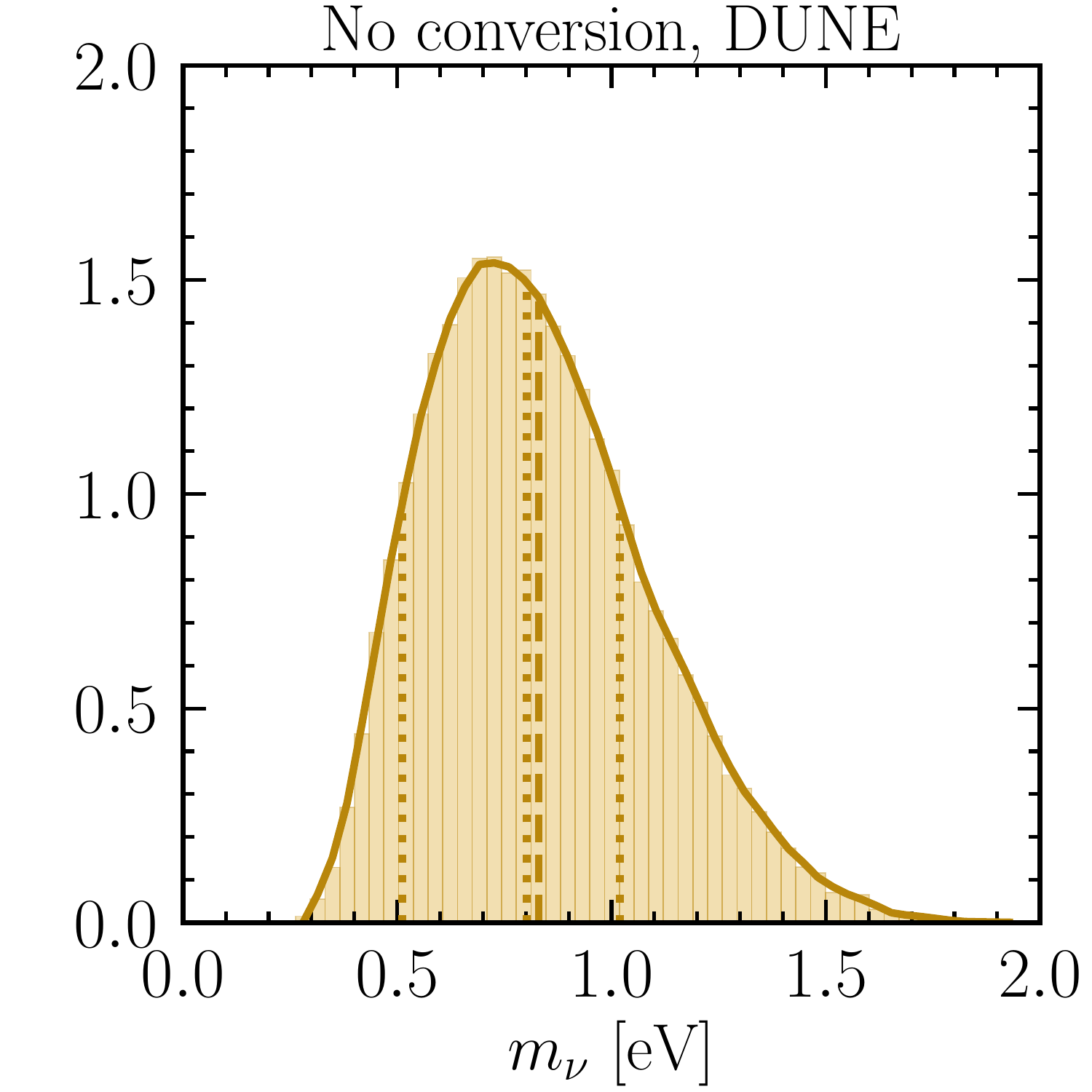} 
\includegraphics[width=0.65\columnwidth]{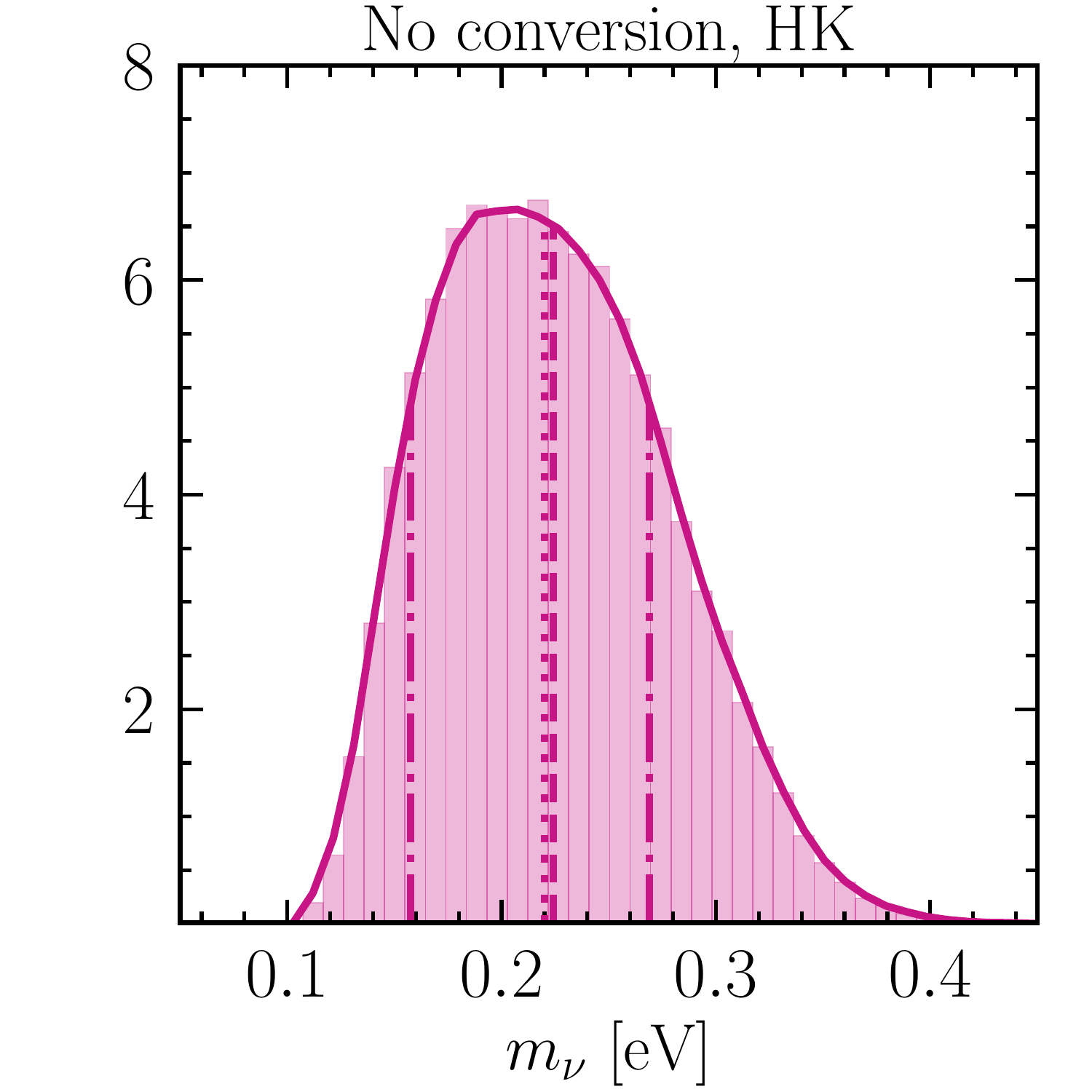} 
\includegraphics[width=0.65\columnwidth]{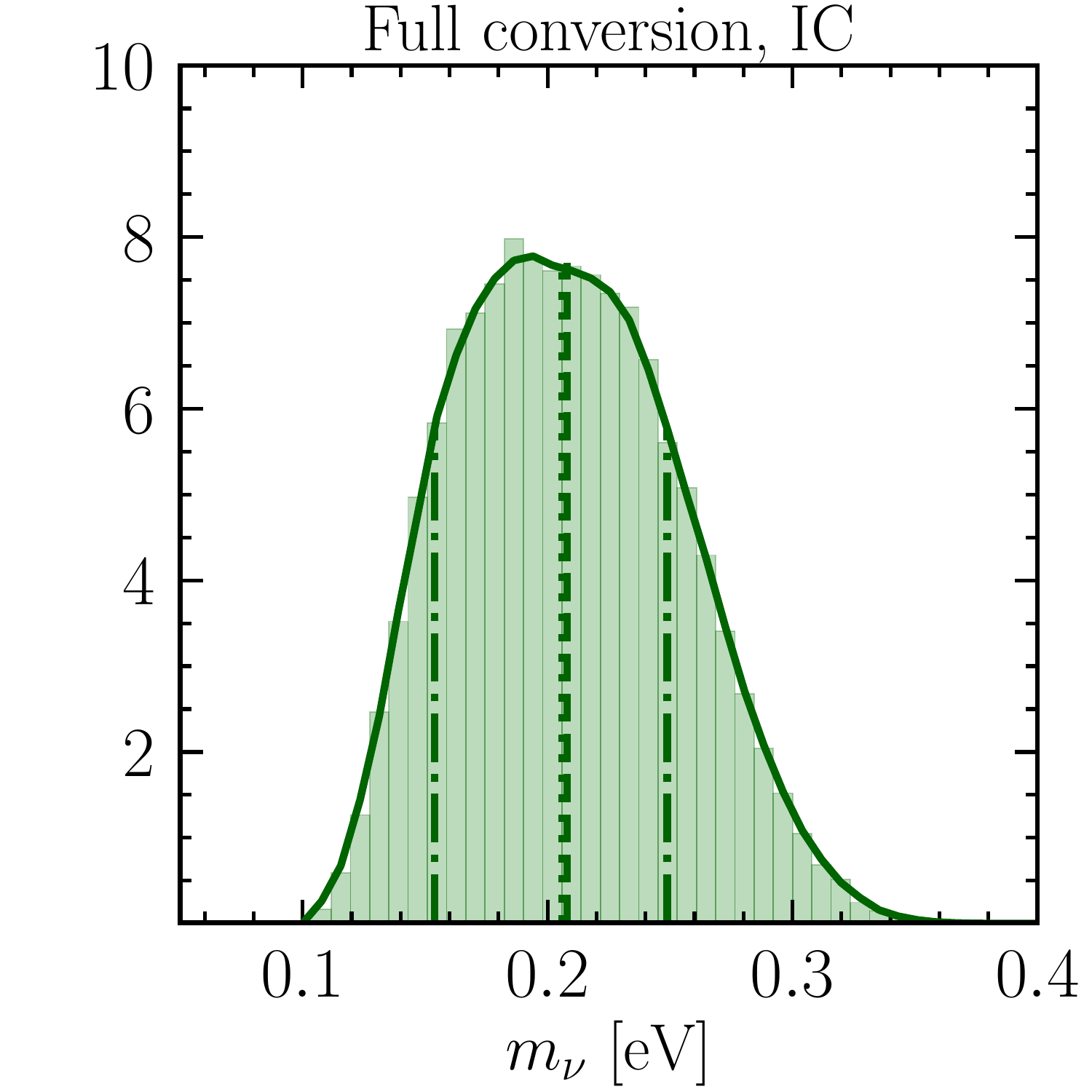}
\includegraphics[width=0.65\columnwidth]{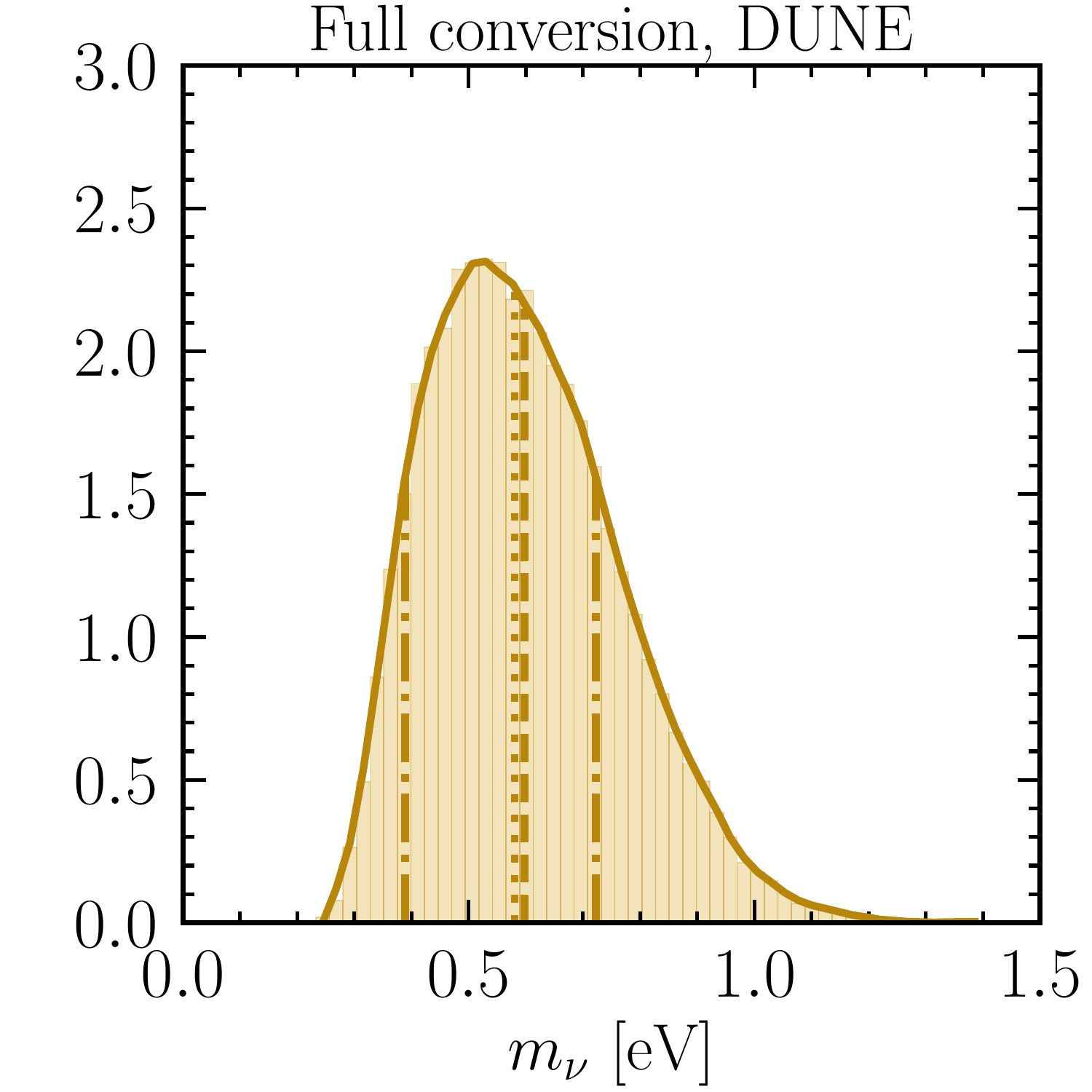} 
\includegraphics[width=0.65\columnwidth]{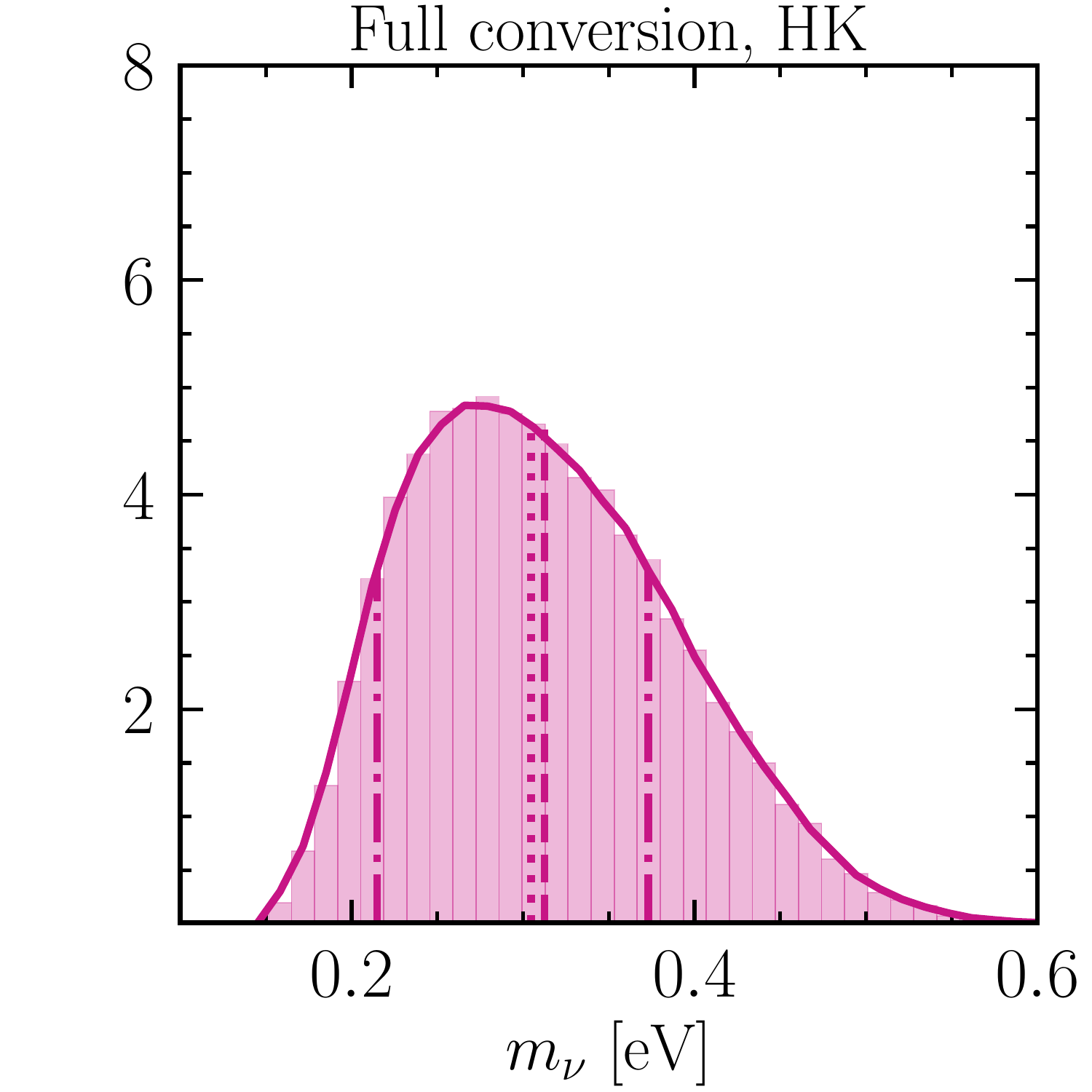} 
\caption{Distributions of the 95~C.L. upper neutrino mass limit for IC (left panels), DUNE (middle panels), and HK (right panels) from the QHPT neutrino burst for a SN at 10~kpc distance; the case of no conversion (upper panels) and full conversion (lower panels) scenarios. Each distribution was constructed with $N_\mathrm{trials} = 5\times 10^4$ and the median, mean, upper uncertainty limit and lower uncertainty limit are plotted with dotted, dashed, and dash-dotted lines. The calculated distributions do not follow the normal distribution.}
\label{Fig:mass_distribiution}
\end{figure*} 

Fig.~\ref{Fig:mass_distribiution} shows the probability density distributions of the upper limits on the absolute neutrino mass calculated using the QHPT peak in the neutrino signal from a SN at a distance 10~kpc from the Earth from IC (left panels; green), DUNE (middle panels; yellow), and HK (right panels; pink) in the no conversion (upper panels) and full conversion oscillation scenarios for $N_\mathrm{trials} = 5\times 10^4$. Although the means (dashed lines) and medians (dotted lines) for each distribution do not differ significantly the skewed high mass tail causes the distribution to diverge from the normal distribution and produces an asymmetric uncertainty band for the calculated mass limits.

\section{Changes of the critical $\chi^2$ value for low neutrino mass}
\label{app:cross-cheks}

\begin{figure*}[t]
\includegraphics[width=0.65\columnwidth]{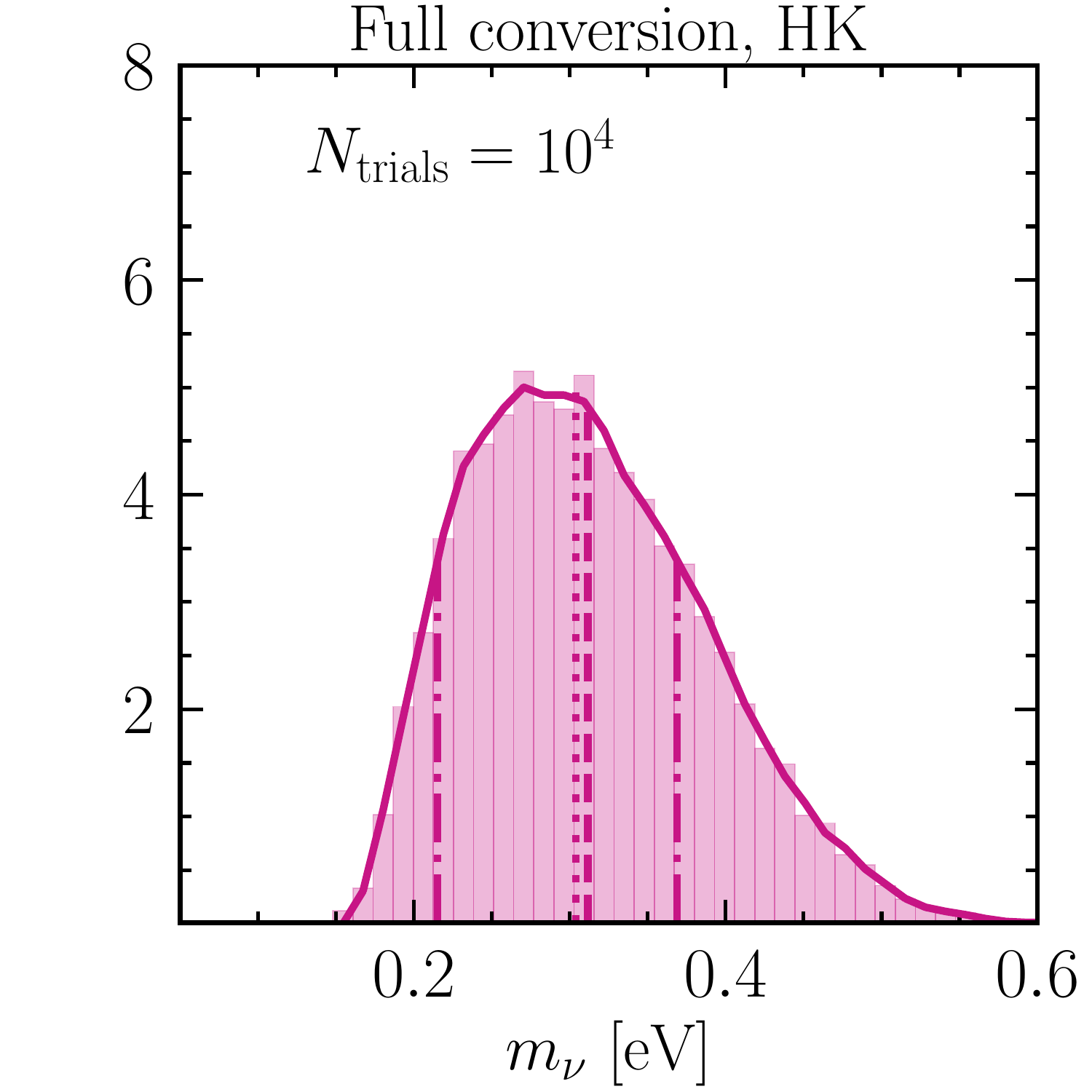}
\includegraphics[width=0.65\columnwidth]{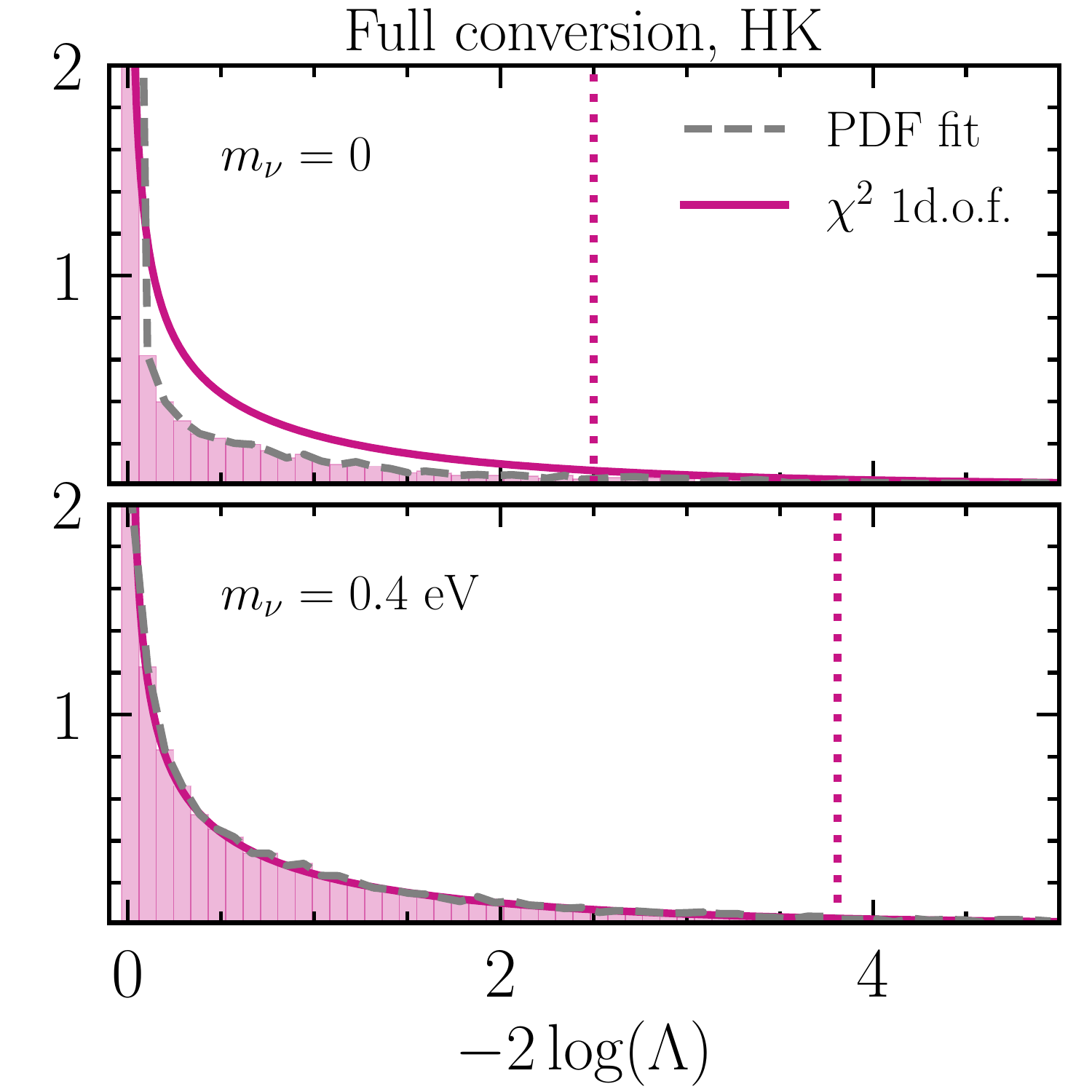}
\includegraphics[width=0.65\columnwidth]{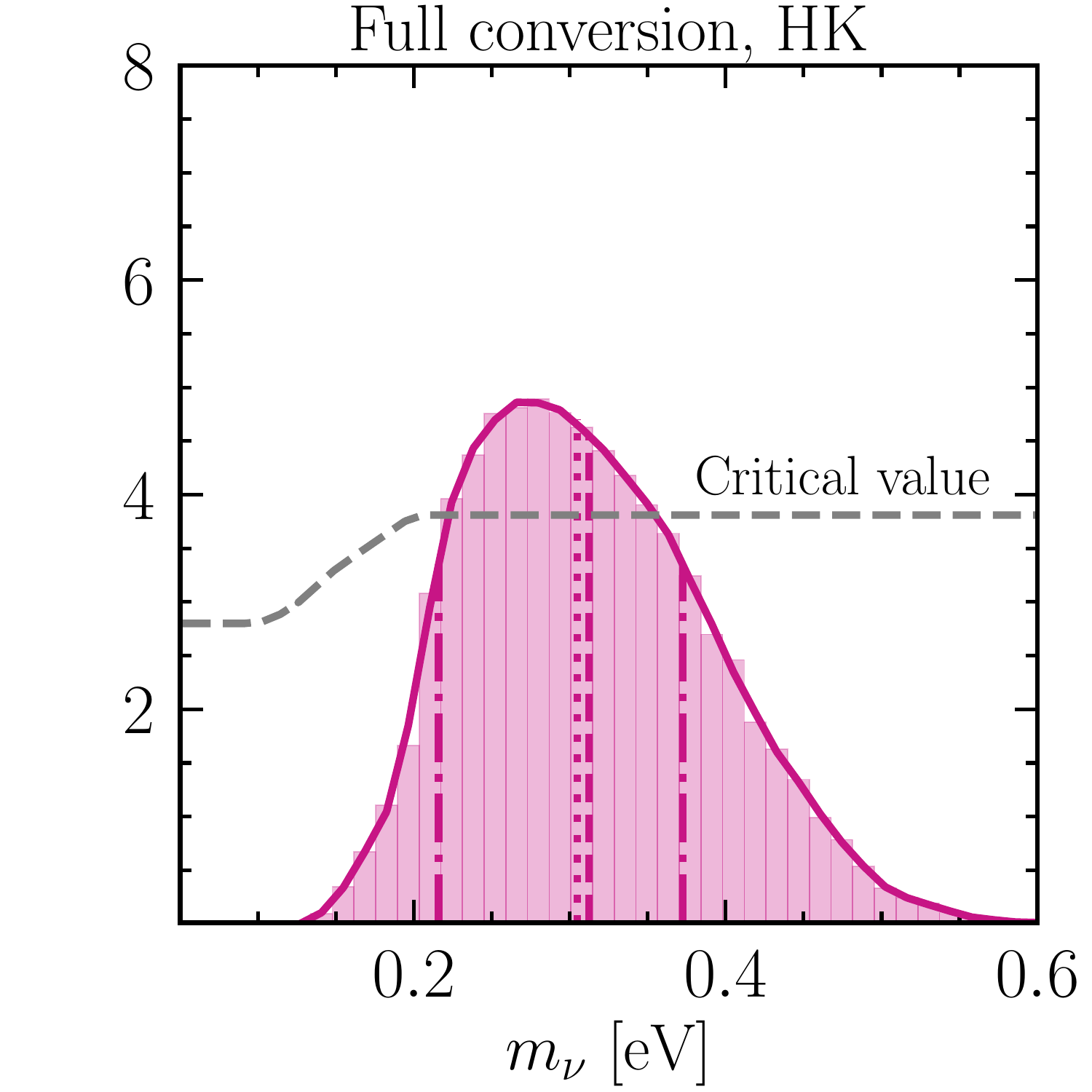}
\caption{\textit{Left panel:} Distributions of the 95~C.L. upper neutrino mass limit for the HK detector from the QHPT neutrino burst for a SN at 10~kpc distance in the "Full conversion" oscillation scenario with $N_\mathrm{trials} = 10^4$.
\textit{Middle panel:} The distribution of the likelihood ratio values for $m_\nu=0$ (upper subpanel) and  $m_\nu=4$~eV (lower subpanel) generated by drawing $10^4$ binned event rates from the corresponding neutrino mass hypotheses. In addition, we show the fits to the obtained distributions (gray), the critical values for the 95\% C.L. upper limits (dotted lines), and $\chi^2$ distribution with 1~d.o.f (solid magenta line). \textit{Right panel:}  Distributions of the 95~C.L. upper neutrino mass limit for the HK detector from the QHPT neutrino burst for a SN at 10~kpc distance in the "Full conversion" oscillation scenario with $N_\mathrm{trials} = 5\times 10^4$ without constructed using the true critical value. The use of the true critical value for the distribution of likelihoods results in slight increase of the number of counts in the left tail of the distribution.}
\label{fig:cross-checks}
\end{figure*} 

The critical value for the hypotheses tested using the likelihood method is often times approximated by the critical value for the $\chi^2$ distribution with the number of d.o.f. corresponding to the number of free parameters of the model, i.e., Wilks' theorem~\cite{Wilks:1938dza}. We have checked that using the Wilks' theorem approximation is valid for sufficiently high masses, whereas for small ones, the critical value obtained from using the Feldman-Cousins prescription~\cite{Feldman:1997qc} is smaller.
The middle panel illustrates the above on the HK "Full conversion" example. Assuming the $m_\nu = 0.4$~eV hypothesis being true, we draw $N_\mathrm{trials} = 10^4$ event rates and calculate the likelihood distribution. It turns out that it closely follow the $\chi^2$ distribution with 1~d.o.f. hence, the critical value of that particular mass is also well approximated by the critical value $\chi^2$ distribution with 1~d.o.f. This is, however, not true for $m_\nu=0$. Due to this effect, the critical value for low masses is reduced. The limits obtained using the true critical value (gray dashed line in right panel of Fig.~\ref{fig:cross-checks}), however, change only slightly, as can be seen in the right panel of Fig.~\ref{fig:cross-checks}. The main difference with respect to distributions calculated assuming Wilks' theorem is a more extended left side tail of the distribution leading to slightly lower limit on the neutrino mass. 
The used approach, Wilks' theorem approximation, therefore, does not make the limits calculated in our work significantly discrepant from the ones obtained by relaxing that assumption.

\bibliography{QGPT}

\end{document}